\documentclass[aps,pra,twocolumn,floatfix,nofootinbib,showpacs,superscriptaddress]{revtex4-2}
\usepackage{mathbbol}
\usepackage{amsmath}
\usepackage{tikz}
\usepackage{centernot}

\usepackage[utf8]{inputenc}
\usepackage[english]{babel}

\usepackage{amsfonts}
\usepackage{amsthm}
\usepackage{mathrsfs}
\usepackage{mathtools} 

\usepackage{physics}
\usepackage{subfigure}

\usepackage[colorlinks=true,hyperfootnotes=true,breaklinks=true,citecolor=red,urlcolor=blue,linkcolor=blue]{hyperref}

\begin{document}
\title{Characterization of partially accessible anisotropic spin chains in the presence of anti-symmetric exchange}
\author{Simone Cavazzoni}
\email{simone.cavazzoni@unimore.it}
\affiliation{Dipartimento di Scienze Fisiche, Informatiche e Matematiche, Universit\`{a} di Modena e Reggio Emilia, I-41125 Modena, Italy}
\author{Marco Adani}
\affiliation{Dipartimento di Scienze Fisiche, Informatiche e Matematiche, Universit\`{a} di Modena e Reggio Emilia, I-41125 Modena, Italy}
\author{Paolo Bordone}
\email{paolo.bordone@unimore.it}
\affiliation{Dipartimento di Scienze Fisiche, Informatiche e Matematiche, Universit\`{a} di Modena e Reggio Emilia, I-41125 Modena, Italy}
\affiliation{Centro S3, CNR-Istituto di Nanoscienze, I-41125 Modena, Italy}
\author{Matteo G. A. Paris}
\email{matteo.paris@fisica.unimi.it}
\affiliation{Quantum Technology Lab, Dipartimento di Fisica {\em Aldo Pontremoli}, Universit\`{a} degli Studi di Milano, I-20133 Milano, Italy}
\affiliation{INFN, Sezione di Milano, I-20133 Milano, Italy}
\date{\today}
\begin{abstract}
We address quantum characterization of anisotropic spin chains in the presence of anti-symmetric exchange, and investigate whether the Hamiltonian parameters of the chain may be estimated with precision approaching the ultimate limit imposed by quantum mechanics.  At variance with previous approaches, we focus on the information that may be extracted by measuring only two neighbouring spins rather than a global observable on the entire chain. We evaluate the Fisher information (FI) of a two-spin magnetization measure, and the corresponding quantum Fisher information (QFI), for 
all the relevant parameters, i.e. the spin coupling, the anisotropy, and the Dzyaloshinskii–Moriya (DM) parameter. Our results show that the reduced system made of two neighbouring spins may be indeed exploited as a probe to characterize global properties of the entire system. In particular, we find that the ratio between the FI  and the QFI is close to unit for a large range of the coupling values. The DM coupling is beneficial for coupling estimation, since it leads to the presence of additional bumps and peaks in the FI and QFI, which are not present in a model that neglects exchange interaction and may be exploited to increase the robustness of the overall estimation procedure. Finally, we address the multiparameter estimation problem, and show that the  model is compatible but sloppy, i.e. both the Uhlmann curvature and the determinant of the QFI matrix vanish. Physically, this means that the state of the system actually depends only on a reduced numbers of combinations of parameters, and not on all of them separately.
\end{abstract}
\maketitle
\section{Introduction}
\label{sec:Introduction}
The coupling constants of an interacting many-body Hamiltonian do not correspond to any observable and one 
has to infer their values by an indirect measurement. In these cases, quantum estimation theory provides 
analytical tools to analyze and optimize the measurement procedure \cite{paris2004quantum,petz2011introduction}. 
In this framework, quantum criticality is considered a resource for estimation, and different theoretical 
models with simple Hamiltonians, such as the Ising model, have been analytically studied to prove 
the generality of this statement \cite{invernizzi2008optimal,zanardi2008quantum,chu2021dynamic}. However, 
more realistic Hamiltonians should consider the presence of anisotropy and of 
Dzyaloshinskii-Moriya (DM) interaction arising from anti-symmetric exchange. Given the evidence 
of phase transitions driven by DM interaction \cite{cepas2008quantum}, phase diagrams of models 
incorporating DM interaction have been investigated \cite{jafari2008phase,jin2017phase}. 
The impact of spin-orbit coupling on crystalline structures \cite{sergienko2006role,halg2014quantum}, interface phenomena \cite{yang2015anatomy}, and spin chains and wires \cite{derzhko2006dynamic,gangadharaiah2008spin,chan2017ising,pylypovskyi2021curvature,fumani2021quantum,pham2021effect} has been studied. Models with DM terms find applications in the computational simulation of realistic systems \cite{di2015direct,dmitrienko2014measuring,yang2023first} and, in the recent years, have been also employed 
to analyze quantum correlations, criticality and factorization of spin chains 
\cite{yi2019criticality,ait2021dynamics}, and are currently attracting attention for applied magnetism \cite{liang2022gradient} and spintronics \cite{kuepferling2023measuring}, with applications in bilayers and multilayer materials \cite{gusev2020manipulation,zhang2022quantifying}, as well as for universal models for quantum computation \cite{wu2002universal} and information \cite{yang2019dynamical}. 

Bipartite entanglement \cite{maruyama2007enhancement,zhang2007thermal,chuan2008entanglement,kargarian2009dzyaloshinskii,park2019thermal}, correlations, and coherence \cite{liu2011quantum,radhakrishnan2017quantum,Haseli_2020} of anisotropic spin chains with anti-symmetric exchange has been analyzed in some details, whereas the precise characterization of the Hamiltonian parameters has been addressed only by global schemes involving the measurement of observables on the entire chain \cite{invernizzi2008optimal,zanardi2008quantum,chu2021dynamic,PhysRevLett.130.240803,Radaelli_2023,PhysRevLett.126.210506}. 
Given the relevance of spin chains in quantum information processing \cite{Lyu_2023,PhysRevA.106.013316}, and the difficulties involved in implementing global observables, we explore here the characterization problem for
partially accessible chains \cite{pa1,pa2,pa3}. In other words, we consider the reduced density matrix 
of two neighbouring spins \cite{lieb1961two,wang2002quantum,wang2002thermal,cai2006robustness,haseli2020entropic},
and investigate whether, and to which extent, information on the value of the chain parameters 
may be extracted by performing measurements only on those two spins. To this aim, we evaluate the QFIs for the chain coupling, the anisotropy, and the 
DM parameter, and the FIs of (two-spin) magnetization measurement. 
Compared to the isotropic case without DM coupling, the FI and the QFI for the coupling constant show additional bumps and peaks, in addition to the local peak related to the phase transition, and we show how this behavior in the presence of DM interaction may be exploited to increase the robustness of the estimation procedure \cite{Ozaydin2015,doi:10.1142/S0217979224503211}. We also analyze the multiparameter case \cite{liu2020quantum}, i.e., the joint estimation of the parameters, with emphasis on compatibility and sloppiness.

The paper is organized as follows. In Section \ref{sec:DM_interaction}, we introduce the theoretical model with its symmetries and describes briefly its main features. In Section \ref{sec:FI_&_QFI}, we introduce classical and quantum Fisher information, and the detailed form of those quantities for an anisotropic XY spin chains with DM interaction. In Section \ref{sec:Coupling_Constant_Estimation}, we present our main analytical and computational findings about the estimation of the coupling constant. Then in Section \ref{sec:P_e_p_t_m_m} we describe in detail how the overall estimation procedure works, and how to exploit the specific features of our system to achieve a metrologically robust and precise estimation scheme. In Section \ref{sec:Multi-parameter_Estimation}, we address the joint estimation of the coupling and the other parameters of the Hamiltonian, showing that the model is compatible, yet sloppy, i.e. the parameters may be in principle jointly estimated without additional quantum noise, but the system actually depends only on a reduced numbers of combinations of parameters, and not on all of them separately. Section \ref{sec:Conclusions} closes the paper with some concluding remarks. 

\section{Planar spin system models with Dzyaloshinsky-Moriya interaction}
\label{sec:DM_interaction}
The Hamiltonian of a $N$ dimensional anisotropic XY spin-half chain in the presence of and external field and of Dzyaloshinskii-Moriya (DM) interaction reads as follows

\begin{align}
	\label{eq:H_syst}
	\mathcal{H} =& \sum_{i=1}^{N} \{J[(1+\gamma)\sigma_{i}^{x}\sigma_{i+1}^{x}+(1-\gamma)\sigma_{i}^{y}\sigma_{i+1}^{y}+ \nonumber \\
       &D(\sigma_{i}^{x}\sigma_{i+1}^{y}-\sigma_{i}^{y}\sigma_{i+1}^{x})]-\sigma_{i}^{z}\},
\end{align}
where $\sigma^{x,y,z}_{i}$ are the Pauli matrices of the i-th spin, $J$ is the coupling constant, $\gamma$ is the anisotropy parameter $\left( -1 \leq \gamma \leq 1 \right)$, and $D$ is the parameter that guides the DM interaction. Notice that the Hamiltonian is expressed in units of the external field (i.e. it is critical for $J=\pm1$), and also that $\hbar = 1$. A fundamental tool for studying the relation between spins of the chain is the reduced density matrix

\begin{equation}
    \label{eq:reduced}
     \rho(i,j)=Tr_{\bar{ij}}(\rho),
\end{equation}
obtained by tracing out all the spins, except for $i$ and $j$, in the total density matrix of the system $\rho$. In principle $\rho(i,j)$ is a $4\times4$ complex hermitian matrix with all the elements different from zero, but due to the symmetries of the Hamiltonian, Eq.\eqref{eq:H_syst}, the reduced density matrix of two spins in the computational basis $\{ \vert 0 \rangle, \vert 1 \rangle \} \bigotimes \{ \vert 0 \rangle, \vert 1 \rangle \}$ has an $X$ structure

\begin{equation}
	\label{eq:reduced DM}
	\rho(i,j) =
\begin{pmatrix}
 	a_+ & 0 & 0 & b_- \\
	0 & c & b_+ & 0 \\
	0 & b_+ & c & 0 \\
	b_- & 0 & 0 & a_-  
\end{pmatrix}.
\end{equation}
The Hamiltonian is translationally ($U(1)$) invariant, so the reduced density matrix $\rho(i,j)$ does not depend separately on the position of the spins $i$ and $j$ but only on their difference $\vert i -j \vert = r $, expressed in terms of reticular sites. Additionally, the system is also invariant upon reflection and so $\rho(i,j) = \rho(j,i)$, and due to the hermiticity of the reduced density matrix $\rho(i,j)=\rho(i,j)^{*}$. The translational invariance and the reflection symmetry implies that the components $\vert 01 \rangle \langle 00 \vert, \vert 10 \rangle \langle 00 \vert, \vert 11 \rangle \langle 01 \vert , \vert 11\rangle \langle 10 \vert$ and their transpose should vanish. Due to these symmetries, the occupation of the state $\vert 01 \rangle \langle 01 \vert$ is equal to that of $\vert 10 \rangle \langle 10 \vert$. The reduced density matrix elements then reads
\begin{equation}
    \begin{split}
    a_{\pm}=&\frac{1}{4}\left( 1 \pm 2 \langle \sigma_i^z \rangle +\langle \sigma_i^z  \sigma_{i+r}^z \rangle \right)\,,  \\
    b_{\pm}=&\frac{1}{4}\left(  \langle \sigma_i^x \sigma_{i+r}^x \rangle \pm \langle \sigma_i^y  \sigma_{i+r}^y \rangle  \right)\,, \\
    c=&\frac{1}{4} \left( 1 - \langle \sigma_i^z \sigma_{i+r}^z \rangle \right)\,.
    \end{split} 
\end{equation}
Assuming to work at zero temperature and in the thermodynamic limit, the magnetization of the spin $i$ is 

\begin{equation}
    \langle \sigma_i^z \rangle =   - \frac{1}{\pi} \int_0^{\pi} d\phi \ \frac{\left[J(\cos \phi - 2D \sin \phi)-1\right]}{\Delta}\, , \forall i\,,
\end{equation}
where the quantity $\Delta$ is given by 
\begin{equation}
    \Delta = \sqrt{\left[ J(\cos \phi - 2D \sin \phi)-1 \right]^2 + J^2 \gamma^2 \sin^2 \phi}.
\end{equation}
The other elements are the correlation functions among the different directions. For neighbouring spins (i.e. $r=1$), we have
\begin{equation}
    \langle \sigma_i^x \sigma_{i+1}^x \rangle = G_{-1}  \,, \forall i 
\end{equation}
\begin{equation}
    \langle \sigma_i^y \sigma_{i+1}^y \rangle = 	G_{1} \, \forall i
\end{equation}
\begin{equation}
    \langle \sigma_i^z \sigma_{i+1}^z \rangle = {\langle {\sigma}_{i}^z \rangle}^2 - G_1 G_{-1} \,, \forall i\,,
\end{equation}
where
\begin{align}
    G_{\pm 1} =& - \frac{1}{\pi} \int_0^{\pi} d\phi \ \frac{2 \cos (\pm \phi )}{\Delta} \left[J(\cos \phi - 2D \sin \phi)-1\right] \nonumber \\
    & + \frac{\gamma}{\pi} \int_0^{\pi} d\phi \ \frac{2 J \sin (\pm \phi )}{\Delta} \sin{\phi} \ .
\end{align}

In the following, we assume that the system is only partially accessible and that only measurements performed on two neighbouring spins are achievable. The above two-spin density matrix is therefore containing the accessible information about the system parameters. 

\section{Fisher and quantum Fisher information}
\label{sec:FI_&_QFI}
In an estimation procedure, the main quantity is the Fisher information (FI)\cite{helstrom1969quantum} of the probability of the outcomes. Starting 
from the case of a single parameter, the FI $F(\lambda)$ of a given 
measurement, say a magnetization measurement, is given by 
\begin{equation}
    \label{eq:general_FI}
	F(\lambda) = \sum_m \frac{1}{p(m|\lambda)} \left(\frac{\partial p(m|\lambda)}{\partial \lambda} \right)^2,
\end{equation}
where ${p(m|\lambda)}=\hbox{Tr}[\rho_\lambda\, \Pi_m]$ is the conditional probability of obtaining the outcome $m$ (an event described by the POVM 
element $\Pi_m$) from a measurement performed on a state labeled by the 
unknown parameter $\lambda$. The Cramer-Rao bound says that the variance $V(\lambda)$ of any (unbiased) estimator of the parameter is bounded by
\begin{equation}
    \label{eq:var}
	V(\lambda) \geq \frac{1}{M\,F(\lambda)},
\end{equation}	
where $M$ is the number of (identically repeated) measurements. 
For the two-spin density matrix of the the previous Section, the FI of 
the two-spin magnetization (i.e. the observable $\sigma_i^z \otimes \sigma_{i+r}^z$) is given by
\begin{equation}
    \label{eq:FI_system}
	F(\lambda)= \frac{1}{a_+} \left(\frac{\partial a_+}{\partial \lambda} \right)^2 + \frac{2}{c} \left(\frac{\partial c}{\partial \lambda} \right)^2 + \frac{1}{a_-} \left(\frac{\partial a_-}{\partial \lambda} \right)^2 \ .
\end{equation}
Upon optimizing over all the possible quantum measurements one obtain that 
$F(\lambda) \leq H(\lambda)$ where the quantum Fisher information 
(QFI) $H(\lambda)$ is defined as
\begin{equation}
    \label{eq:general_qfi}
    H(\lambda) = \max_{\{\Pi\}} F(\lambda) 
    =\hbox{Tr}[\rho_{\lambda}\mathcal{L}_{\lambda}],
\end{equation}
where $\mathcal{L}$ is the so-called symmetric logarithmic derivative, 
defined by the implicit relation
\begin{equation}
\label{eq:sld}
\partial_{\lambda} \rho_{\lambda} = \frac{1}{2} \{ \mathcal{L}_{\lambda}, \rho_{\lambda} \} = \frac{1}{2}\left( \mathcal{L}_{\lambda}\rho_{\lambda} + \rho_{\lambda}\mathcal{L}_{\lambda} \right).
\end{equation}
Overall, we have  that the variance of any estimator is bounded by
\begin{equation}
    \label{eq:qcr}
	V(\lambda) \geq \frac{1}{M\,F(\lambda)} \geq \frac{1}{M\,H(\lambda)},
\end{equation}

For $X$ states is always possible to decompose the density matrix in the sum of two commuting matrices as $\rho_\lambda = \rho_{1\lambda} +\rho_{2\lambda}$, where in our case
\begin{equation}
	\label{eq:rho_1}
	\rho_{1\lambda} =
\begin{pmatrix}
 	a_+ & 0 & 0 & b_- \\
	0 & 0 & 0 & 0 \\
	0 & 0 & 0 & 0 \\
	b_- & 0 & 0 & a_-  
\end{pmatrix},
\end{equation}
and 
\begin{equation}
	\label{eq:rho_2}
	\rho_{2\lambda} =
\begin{pmatrix}
 	0 & 0 & 0 & 0 \\
	0 & c & b_+ & 0 \\
	0 & b_+ & c & 0 \\
	0 & 0 & 0 & 0  
\end{pmatrix}.
\end{equation}
Doing so, also the QFI $H(\lambda)$ can be decomposed in the sum of the QFI associated to the two matrices \cite{maroufi2021analytical}, as
\begin{equation}
	H(\lambda) = H_1(\lambda) + H_2(\lambda) \ .
\end{equation}
where the two QFIs may be written as
\begin{align}
    \label{eq:H_rho1}
    H_1(\lambda) = & \frac{1}{\omega_0} \left[ \frac{{(g_{\alpha \beta} \omega^\alpha \partial_\lambda \omega^\beta)}^2}{g_{\alpha \beta} \omega^\alpha  \omega^\beta} - g_{\alpha \beta} \left( \partial_\lambda \omega^\alpha \right) \left(\partial_\lambda \omega^\beta \right) \right] \nonumber \\ & 
    + \frac{(\partial_\lambda \omega_0)^2}{\omega_0} \ ,
\end{align}
and
\begin{align}
    \label{eq:H_rho2}
        H_2(\lambda) =&\frac{1}{\tilde{\omega}_0} \left[ \frac{{(g_{\alpha \beta} \tilde{\omega}^\alpha \partial_\lambda \tilde{\omega}^\beta)}^2}{g_{\alpha \beta} \tilde{\omega}^\alpha  \tilde{\omega}^\beta} - g_{\alpha \beta} \left( \partial_\lambda \tilde{\omega}^\alpha \right) \left(\partial_\lambda \tilde{\omega}^\beta \right) \right] \nonumber \\ 
        & + \frac{(\partial_\lambda \tilde{\omega}_0)^2}{\tilde{\omega}_0}\ .
\end{align}
The $\omega_\alpha$s and the ${\tilde{\omega}}^{\alpha}$s, with $\alpha = 0, 1, 2, 3$ are
given by
\begin{equation}
    \begin{split}
    	\omega_0 & = \frac{1}{2}\left( 1 + \langle  \sigma_i^z  \sigma_{i+r}^z \rangle \right),  \ \ \ \ \omega_3 =  \langle \sigma^z \rangle, \\  \omega_1 & =\frac{1}{2}\left( \langle  \sigma_i^x  \sigma_{i+r}^x \rangle - \langle  \sigma_i^y  \sigma_{i+r}^y \rangle \right),  \ \ \ \ \omega_2 = 0 \,
    \end{split}
\end{equation}

\begin{equation}
    \begin{split}
		\tilde{\omega_0} & = \frac{1}{2}\left( 1 - \langle  \sigma_i^z  \sigma_{i+r}^z \rangle \right),  \ \ \ \ \tilde{\omega_3} =  0 \ ,\\  \tilde{\omega_1} & =\frac{1}{2}\left( \langle  \sigma_i^x  \sigma_{i+r}^x \rangle + \langle  \sigma_i^y  \sigma_{i+r}^y \rangle \right),  \ \ \ \ \tilde{\omega_2} = 0 \ .
   \end{split}
\end{equation}
and $g_{\alpha \beta} = diag\{1,-1,-1,-1\}$ is the Minkowski metric introduced to simplify the notation.

If the estimation procedure involves more than one parameter, i.e. $\pmb{\lambda} \in \mathbb{R}^{n}$, the FI and QFI become symmetric positive definite matrices \cite{albarelli2020perspective}, whose elements are defined as
\begin{equation}
    \label{eq:FI_multiparameter}
    F_{\mu \nu}  = \sum_{k} p(k \vert \pmb{\lambda}) [\partial_{\mu} \log p(k \vert \pmb{\lambda})] [\partial_{\nu} \log p(k \vert \pmb{\lambda})] ,
\end{equation}
and
\begin{equation}
    \label{eq:QFI_multiparameter}
    H_{\mu \nu}  = \hbox{Tr} \left[ \rho_{\pmb{\lambda}} \frac{ \mathcal{L}_{\mu} \mathcal{L}_{\nu} + \mathcal{L}_{\nu}\mathcal{L}_{\mu}}{2} \right]. 
\end{equation}
The Cramer-Rao bound becomes a matrix relation for the covariance matrix of any set of estimators
\begin{equation}
    \label{eq:var_multiparameter}
        \textbf{V}(\pmb{\lambda}) \geq \frac{1}{M} \textbf{F}^{-1}(\pmb{\lambda}) 
        \geq \frac{1}{M} \textbf{H}^{-1}(\pmb{\lambda}).
\end{equation}
In the joint estimation of two parameters $\lambda_{\mu},\lambda_{\nu}$, some additional intrinsic noise of quantum origin is present if the two symmetric logarithmic derivatives do not commute \cite{carollo2019quantumness,raze20}
\begin{equation}
    \label{eq:comm_L}
    \left[ \mathcal{L}_{\nu},\mathcal{L}_{\mu} \right] \neq 0 \ .
\end{equation}
The information about the commutativity between all the pairs of symmetric logarithmic derivatives, is provided by the Uhlmann matrix, with elements  defined as
\begin{equation}
    \label{eq:U_multiparameter}
    U_{\mu \nu}  = \hbox{Tr} \left[ \rho_{\pmb{\lambda}} \frac{ \mathcal{L}_{\mu} \mathcal{L}_{\nu} - \mathcal{L}_{\nu}\mathcal{L}_{\mu}}{2} \right] \ . 
\end{equation}
The highest is the value of the elements $\abs{U_{\mu \nu}}$ the more incompatible is the joint estimation procedure of the two parameters $\lambda_{\mu}$ and $\lambda_{\nu}$. If the elements of the Uhlmann matrix are vanishing $U_{\mu \nu}=0$ the measurements are compatible and no intrinsic noise affects the joint estimation.

\section{Estimation of the coupling constant $J$}
\label{sec:Coupling_Constant_Estimation}
Among the parameters of the Hamiltonian, the coupling constant $J$ is the 
one of main interest. It does not correspond to a physical observable, and should be estimated from the measurement of another quantity.  It is well known 
that when the external field equals the value of coupling constant 
($J=\pm1$ for our renormalized Hamiltonian)  
the system shows a phase transition \cite{dziarmaga2005dynamics}, which can 
be exploited as a resource in quantum metrology .
\subsection{Anisotropic Heisenberg XY spin chain}
We start our analysis by looking at the QFI $H(J)$, bounding the 
precision in the estimation of the coupling constant $J$, in the 
case $D=0$ and look at the dependence on the anisotropy parameter 
$\gamma$. For $D=0$, the model reduces to the Heisenberg XY spin 
chain, in which the components of the spins interacts differently 
in $x$ and $y$ direction due to the effect of the anisotropy 
$\gamma$. 
\begin{figure}[htb]
    \includegraphics[width=0.9\columnwidth]{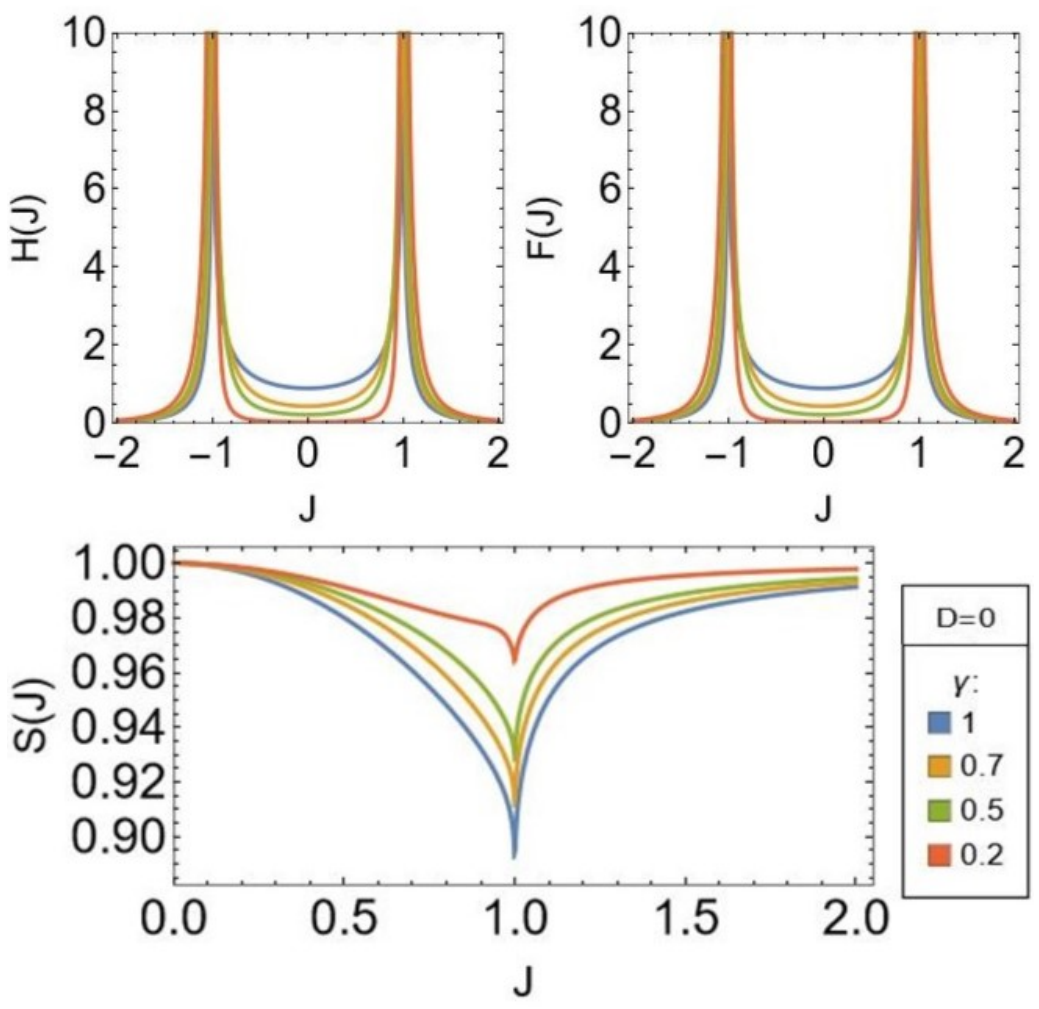}
    \caption{Top left panel: QFI $H(J)$ for the two-spin reduced density 
    matrix of an anisotropic Heisenberg XY spin chain, i.e. for $D=0$. Top right panel: magnetization FI. Bottom panel: the saturation ratio $S(J)=F(J)/H(J)$.  The different colours of the curves correspond to different values of the anisotropy parameter $\gamma = \{0.2, 0.5, 0.7, 1\}$ (see legend), 
    $\gamma = 1$ corresponds 
to the Ising Model.}
    \label{fig:collage_D=0.jpg}
\end{figure}
Looking at Fig.\ref{fig:collage_D=0.jpg}, we first notice that for $D=0$ 
all the curves are even independently of $\gamma$ (i.e., $H(J)=H(-J)$ 
$\forall \gamma$). The Hamiltonian is symmetric with respect to $\gamma$. 
The effect of the transformation $\gamma \rightarrow -\gamma$ does not 
affect the QFI, since it only exchanges the $x$ 
and $y$ components of the spins. We can then focus only on positive 
values of $\gamma$ in Fig.\ref{fig:collage_D=0.jpg}. 

The figure shows that by measuring only two neighbouring spins is 
enough to understand the collective behavior of the system. The QFI 
shows a sharp peak for $J=\pm1$, thus sensing the phase transition 
between ferromagnetic and antiferromagnetic regime. As $J$ approaches 
the values $J=\pm2$, the value of $H(J)$ is consistently low and 
keeps decreasing as the value of $J$ moves away from $J=\pm1$. 
Moving from $J=0$ to $J=\pm1$, the curves have greater values if 
the anisotropy is higher, yet near the two divergences, this behavior 
is reversed, and the higher curves are those corresponding to lower 
$\gamma$. 

Upon approaching the divergences from the paramagnetic region 
($J<-1$ or $J>1$), the higher QFI is observed for lower anisotropy. 
Far from the divergences, the highest value of the QFI is reached for 
$\gamma=1$ (i.e., for the Ising Model), while near $J=\pm1$ the 
divergence is more pronounced for lower anisotropy. In the limit 
$\gamma \rightarrow 0$, the QFI vanishes, i.e. the isotropic Heisenberg 
XY spin chain is definitely not suitable to precisely characterize the 
coupling. 

The behavior of the magnetization FI is very 
close to that of the QFI, showing that the ultimate quantum bound to 
precision may be achieved by a feasible measurement. A quasi local 
magnetization measurement involving only two neighbouring spins is 
indeed capable of capturing the collective behavior of the system. 
The FI shows the same symmetries for $J\rightarrow-J$, 
and $\gamma \rightarrow -\gamma$, the same dependence on the anisotropy 
parameter $\gamma$, and notably the same divergence near the phase 
transitions. To better quantify the effectiveness of the magnetization measurement, we introduce the ratio between $F(J)$ and $H(J)$. This 
quantity, defined as
\begin{equation}
    \label{eq:saturation_S}
    S(J)=\frac{F(J)}{H(J)},
\end{equation}
is referred to as \textit{saturation} since shows how much the 
inequality $F\leq H$ saturates to an equality. From 
Fig.\ref{fig:collage_D=0.jpg}, we see that the \textit{saturation} 
is always considerably high (above 0.89) for all the considered 
combinations of Hamiltonian parameters. Starting from $J=0$, the 
saturation is equal to 1, then moving to $J=1$, it decreases until 
it reaches its minimum. From $J=1$ to $J=2$ the saturation increases 
again and all the curves are above $S=0.98$. We can notice that for 
the cases we have analyzed, i.e. $\gamma={0.2, 0.5, 0.7, 1.0}$, the 
lower is the anisotropy the higher are the curves $S(J)$. This means 
that the fraction of the information that is possible to extract 
from the system through magnetization is higher for lower 
anisotropy.
\subsection{Effects of the DM interaction} 
\begin{figure}[h!]
    \centering
    \includegraphics[width=0.9\columnwidth]{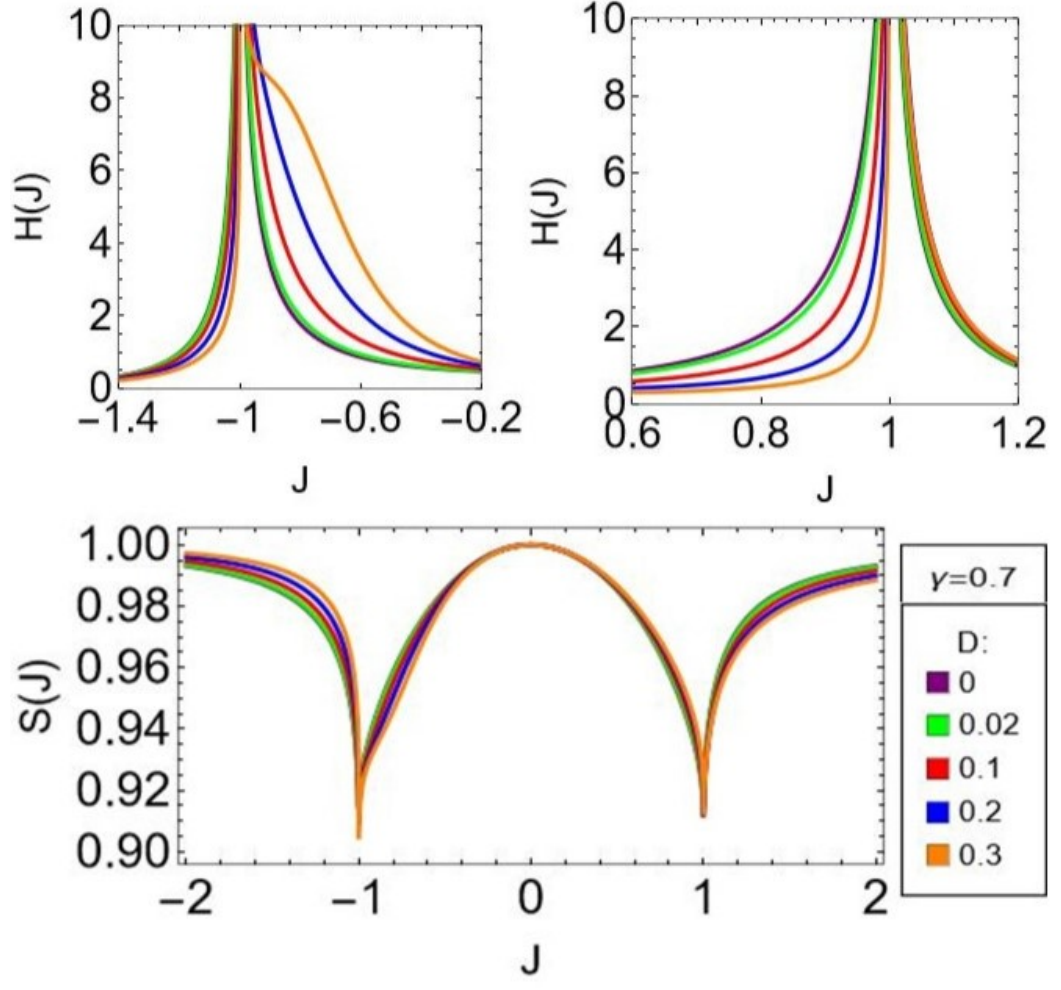}
    \caption{Top panels: TheQFI $H(J)$ of the two-spin reduced density matrix of an anisotropy Heisenberg XY spin chain ($\gamma=0.7$), and different values of the  DM coupling. On the top left panel: the region in proximity of the $J=-1$ divergence. On the right: the region in vicinity of the $J=1$ divergence. Bottom panel: the saturation function $S(J)$. The different curves correspond to different values of the parameter $D = \{0, 0.02,0.1,0.2,0.3\}$.} 
    \label{fig:collage_Dnot_g0.7.jpg}
\end{figure}

\begin{figure}[htb]
    \centering
    \includegraphics[width=0.9\columnwidth]{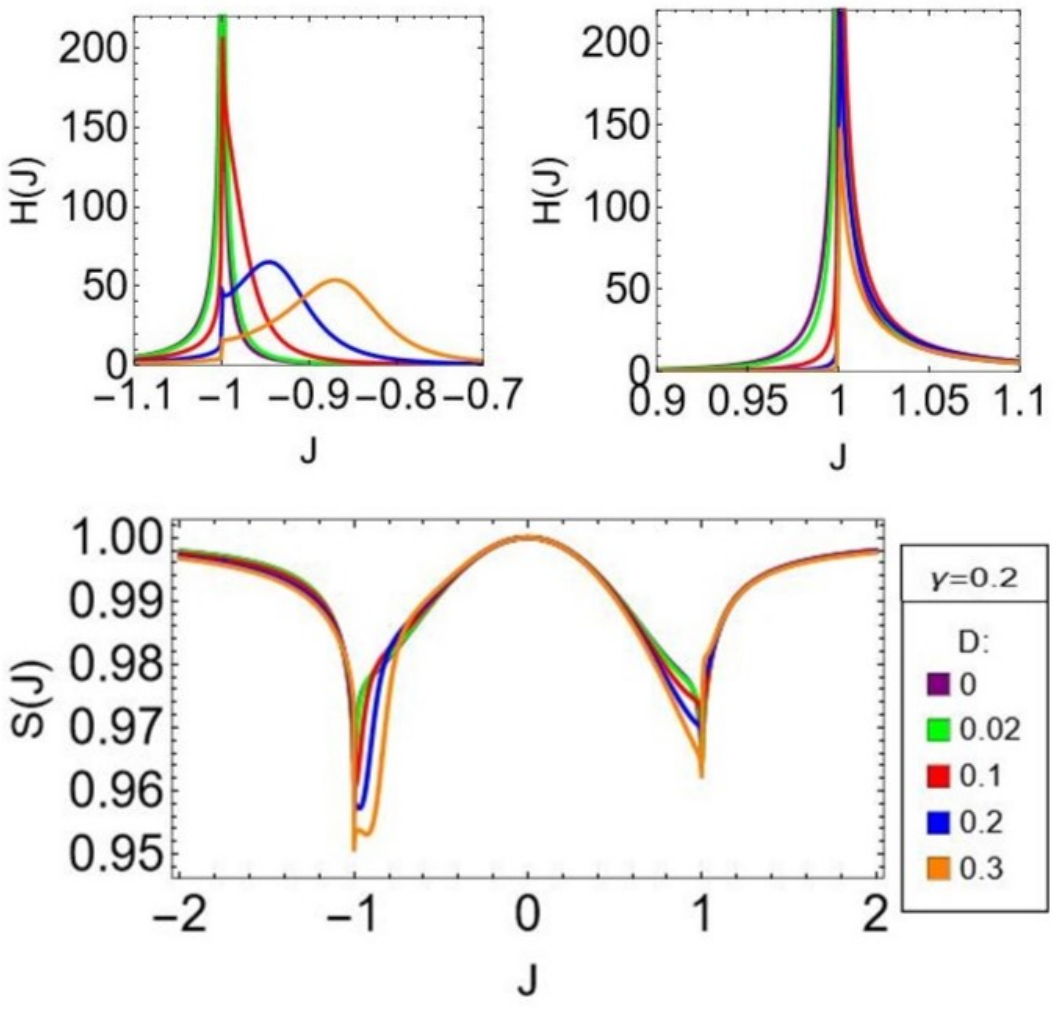}
    \caption{Top panels: The QFI $H(J)$ for a two-spin reduced density matrix of an anisotropy Heisenberg XY spin chain ($\gamma=0.2$), and different values of the  DM coupling. On the top left panel: the region in proximity of the $J=-1$ divergence. On the right: the region in vicinity of the $J=1$ divergence. Bottom panel: the saturation function $S(J)$. The different curves correspond to different values of the parameter $D = \{0, 0.02,0.1,0.2,0.3\}$.}
    \label{fig:collage_Dnot_g0.2.jpg}
\end{figure}

Once the effect of the anisotropy parameter $\gamma$ has been clarified, 
let us move to the study of the effects of the DM interaction on the 
estimation of $J$. Results are illustrated in Figs.\ref{fig:collage_Dnot_g0.7.jpg} and \ref{fig:collage_Dnot_g0.2.jpg}. 

The first relevant effect of $D$ is to break the symmetry of $H(J)$, which is no more an even function of $J$. On the other hand, we have a novel symmetry, due to the presence 
of the product $J\cdot D$ in the Hamiltonian. As a consequence,  we focus to the case $-2 \leq J \leq 2$ and $D \geq 0$, because for $D<0$ the results can be obtained through a reflection across the line $J=0$. At the left of the negative divergence $(-2 \leq J < -1)$ the curves are lower for higher $D$. In the region on the right, $-1 < J < 0$, close to the divergence the curves are the higher the lower is $D$. Moving away from $J=-1$ we observe different behaviors in different ranges of $D$. From $D=0$ to a threshold value that we call $D_{bump}(\gamma)$, the curves are the higher the higher is $D$. Then up to another threshold value called $D_{peak}(\gamma)$, so for $D_{bump}(\gamma)<D<D_{peak}(\gamma)$, a bump appears in the curves, as it happens for the curve associated to $D=0.3$ in Fig.\ref{fig:collage_Dnot_g0.7.jpg} and for the curve associated to $D=0.1$ in Fig.\ref{fig:collage_Dnot_g0.2.jpg}. In the curve associated to $D=D_{peak}(\gamma)$ the bump becomes an inflection point. For $D>D_{peak}(\gamma)$ a peak appears in the curves, as we can see it in the curves associated to $D=0.2$ and to $D=0.3$ in Fig.\ref{fig:collage_Dnot_g0.2.jpg}. For positive $J$, Fig.\ref{fig:collage_Dnot_g0.7.jpg} and Fig.\ref{fig:collage_Dnot_g0.2.jpg}, in the region on the left of the $J=1$ divergence, (i.e. $0 < J < 1$), the curves are the higher the lower is $D$, as in the region $-2 \leq J < -1$. In the interval $1 < J \leq 2 $ near to the divergence again the curves are the higher the lower is $D$. Getting away from $J=1$, the curves cross each other, then, for $D$ lower than a threshold value called $D_{loss}(\gamma)$ the higher is $D$ the higher are the curves. When $D>D_{loss}(\gamma)$ the curves start to became lower than the one for $D=D_{loss}(\gamma)$ at the beginning and at the end of the interval. Keeping to increase $D$ the curves become lower than the one for $D=D_{loss}(\gamma)$ in all the interval. These features may be exploited to make the overall estimation procedure more robust, as it will discussed in the next Section.

As in the case $D=0$, also in the presence of DM interaction the magnetization FI is close to the QFI. A two-spin magnetization measurement is thus able to achieve the ultimate precision.  Looking at the saturation $S(J)$ in Fig.\ref{fig:collage_Dnot_g0.7.jpg} and Fig.\ref{fig:collage_Dnot_g0.2.jpg} we see 
how the FI  is quantitatively close to the QFI for $\gamma=0.7$ and $\gamma=0.2$.

In the first case, the saturation $S(J)$ is always above 0.9. In the second case it is always above 0.95. These results tell that the measurement procedure under analysis works properly also in the presence of DM coupling. Moreover, the new features introduced by $D \neq 0$ (bump or peak) can be exploited to increase the robustness of the estimation procedure. In both cases, as the value of $D$ increases, the height of the main peak in $J=-1$ decreases. At the cost of a reduction in precision, therefore there is a gain in \textit{robustness} of the measurement, which means that even if the maximum value of the QFI diminishes, it has an appreciable value in a larger range of $J$.

\section{Practical estimation procedure through magnetization measurements}
\label{sec:P_e_p_t_m_m}
To fully understand the practical implications of our results, it is necessary to understand how the overall estimation procedure of the coupling constant works. This procedure requires the possibility to control the external magnetic field $B$. This is crucial, since varying the value of $B$ we can exploit the peaks in the FI as a metrological resource. So it is possible, at least in principle, to enhance the precision for the estimation of the parameter $J$ regardless its value. Another relevant characteristic of the estimation procedure is the relative sign between $J$ and $B$, because it is possible to exploit the different behaviors of  $F(J/B)$ when $J/B>0$ or $J/B<0$ to obtain a more robust or precise measurement. 
\paragraph{General notions ---} 
The estimation procedure works as follows. We denote the true value of $J$ of the system $J_{s}$. In the estimation procedure, we have to start from an initial guess, that we call $J_{guess}$. From this, we set the external magnetic field $B=J_{guess}$. Because of the main peak of the FI, the closer is $J_{guess}$ to $J_{s}$ the lower is the variance of any estimator. After setting $B=J_{guess}$, we perform a set of magnetization measurements. Then, we map the set of outcome to an estimate the parameter (i.e. we use an \textit{estimator}\cite{paris2009quantum}) and the value $J_{av,1}$, with its associated variance. To improve the precision of this result, we can set $B=J_{av,1}$ and repeat the measurements to find a new estimate for $J$, again with its associated variance. As before, the external magnetic field has to be re-set to the new average value of $J$, $B=J_{av,2}$. Going on with this procedure $n$ times we have  $J_{av,1} \rightarrow  J_{av,2}  \rightarrow  ... \rightarrow J_{av,n}$. If variance decreases step by step, i.e., $V(J_{av,1}) > V(J_{av,2}) > ... > V(J_{av,n})$, the  $J_{av,n}$ becomes closer and closer to $J_{s}$, and the procedure converges. This happens because when $J_{av,n}$ approaches $J_{s}$, the closer $J_{s}$ 
is to the main peak, the larger is $F(J_{s})$ and the lower is the related variance. 

If the initial guess $J_{guess}$ is too different from $J_{s}$, it could happen that the variance 
does not decrease step by step. In this case, we cannot ensure that the procedure converges 
to $J_{s}$ and could be better try to modify the initial guess. For this reason the higher is the FI of the coupling constant, associated to the interval of values of $J$ that is used in the estimation, the more probable is to obtain a value of $J_{av,n}$ that is closer to $J_{s}$ even if the tuning between the external field $B$ and $J_{s}$ is not yet accurate. This means that the higher is $F(J)$ in the working interval of $J$ the higher is the robustness of the procedure.

\paragraph{Estimation  without DM interaction ---} 
As described in Section \ref{sec:Coupling_Constant_Estimation}-A, $F(J)$ is even in $J$, so in this case the relative sign between $J_{s}$ and $B$ is not relevant. Looking at the behavior of the FI curves in Fig.\ref{fig:collage_D=0.jpg}, we notice how the robustness of the estimation procedure is higher for higher $\gamma$ in interval $0 < J < 1$, while it is the opposite for the interval $1 < J \leq 2$. So for example if $\gamma = 1$, it is convenient to overestimate $J_{guess}$ compared to underestimate it, whereas for $\gamma=0.2$ is the opposite. We see that in the optimal conditions, the robustness of the estimation is higher for larger $\gamma$, while the convergence speed close to the divergence is higher for 
lower $\gamma$. 

\paragraph{Estimation in the presence of DM interaction ---} 
As described in Section \ref{sec:Coupling_Constant_Estimation}-B, when $D\neq0$, $F(J)$ is no more even in $J$, so the behavior is different for $J<0$ or $J>0$. For this reason the relative sign between $B$ and $J_{s}$ matters and can be used to select the region of $F(J)$ for $J$ positive or negative, to work with. In both regions, in the intervals on the left of the two divergences (i.e. $-2 \leq J < -1$ and $0 < J < 1$) $F(J)$ is lower then in the regions on the right of the divergences (i.e. $-1 < J <0$ and $1 < J < 2$). This means that the overestimation of $J_{guess}$ is convenient respect the underestimation for $J<0$ while for $J>0$ is the opposite. In the interval $-1<J<0$ the sub-interval associated to the values of $F(J)$ that are significantly different from 0 is wider for higher $D$, and it is always wider then the analogous sub-interval in the region $1 < J \leq 2$. This implies that the robustness of the estimation procedure is higher working in the region $J<0$ of $F(J)$. In particular in this region the robustness is always higher for $D\neq 0$. On the other hand, from Fig.\ref{fig:collage_Dnot_g0.7.jpg} is clear how much $H(J)$ (and consequently $F(J)$ due to the behavior of the \textit{saturation} parameter) decreases as $D$ increases near $J=-B$. This loss in the values of $F(J)$ is not equally relevant near the $J=B$ divergence. 

Overall, we conclude that the behavior of $F(J)$ may be successfully and effectively exploited to make the estimation procedure is more robust respect to the case $D=0$. Moreover, when we are sufficiently close to $J_{s}$, we may implement a change of the relative sign between $B$ and $J_{s}$ to switch to the region $J>0$. In this way, we can exploit the higher values of $F(J)$ close to $J=B$ to complete the refinement of the estimation.  

\section{Multiparameter estimation}
\label{sec:Multi-parameter_Estimation}
After the estimation of the coupling constant $J$ we move forward to the 
multiparameter estimation case. In principle, all the quantities in the 
Hamiltonian may not be known \textit{a priori}, and then, also the DM 
interaction parameter $D$ and the anisotropy parameter $\gamma$ have to 
be estimated. In this Section we focus on two examples of multiparameter 
estimation. In the first, we assume to have a large amount of {\em a priori} 
information about the coupling and study the dependence of the QFI matrix (QFIM) on the 
DM parameter and the anisotropy. In the second example, we assume to have 
a large amount of {\em a priori} 
information about the anisotropy and study the dependence of the QFIM on the 
coupling and the DM parameter.

\subsection{Joint estimation at fixed coupling}
As a first example, we assume that the coupling constant have been previously estimated 
with a sufficient high precision (here we set $J=0.999$) and address
the joint estimation of the three parameters, looking at the properties 
of the QFI matrix as a function of $D$ for different values of $\gamma$. 

\begin{figure}[htb]
    \centering
    \includegraphics[width=0.9\columnwidth]{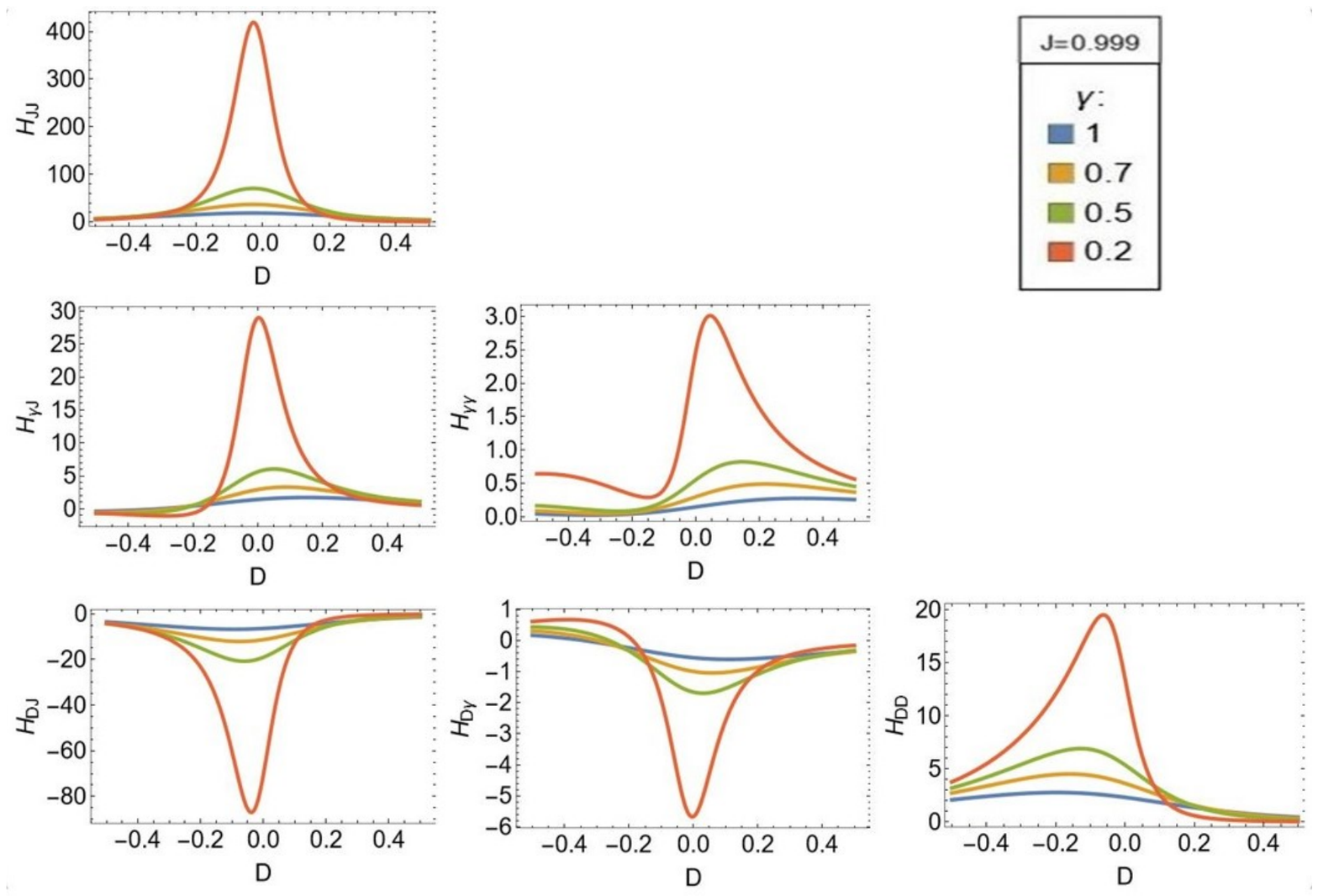}
    \caption{The matrix elements of the  QFIM $H(J,\gamma,D)$ for a two-spin reduced density matrix of an anisotropy spin chain with $D\neq 0$ as a function of $D$, and different values of $\gamma$ (the coupling is set to $J=0.999$). The array of plots corresponds to the position of the elements $H_{\mu \nu}$ in the QFIM, and the different curves in each plot denote the results for different values of the anisotropy parameter $\gamma$ (see the legend).}
    \label{fig:qfi_matrix_J=0.999}
\end{figure}

As shown in Fig.\ref{fig:qfi_matrix_J=0.999} the diagonal element of the QFI $H_{JJ}(D)$, in the region close to $D=0$, clearly shows a maximum, independently on the value of $\gamma$, and symmetrically decreases as $D$ increases or decreases. The other two diagonal elements, $H_{\gamma \gamma}(D)$ and $H_{DD}(D)$ show almost a complementary behavior independently on the value of the anisotropy parameter. $H_{\gamma \gamma}$ is very small for negative $D$, then increases rapidly, shows a maximum around $D=0$ and then decreases again. On the other hand, $H_{DD}$ is considerably high for negative $D$ (value of $D$ opposite to $J$), shows again a maximum around $D=0$, and then decrease up to approximately 0 as $D$ increases. 

As an overall effect, the parameter $\gamma$ worsens the estimation procedure. The main difference among the diagonal elements is in their absolute values: $H_{JJ}$ varies between 0 and 400, $H_{\gamma \gamma}$ between 0 and 3, and $H_{DD}$ between 0 and 20, which means that a very different number of measurements are needed to estimate these three parameters with the same precision. The QFI of the coupling constant $J$ as a function of $D$ shows  maximum around the value $D=0$, as observed also in the single parameter estimation 
(see Section \ref{sec:Coupling_Constant_Estimation}) for increasing $\gamma$ this 
maximum value decreases. Regarding the estimation of the $DM$ parameter $D$, the element $H_{DD}(D)$, is large when $D$ has the opposite sign of $J$. When $J$ and $D$ have the same sign $H_{DD}(D>0)$ is almost zero, while the behavior is reversed for the anisotropy parameter $\gamma$, i.e., $H_{\gamma \gamma}(D)$ is low for negative $D$ and increases with $D$. The main difference among all the three elements is the values they assume in the range $-0.4<D<0.4$. As the anisotropy parameter $\gamma$ increases, the dependence on $D$ becomes less and less evident for all the diagonal components of the QFIM. $H_{JJ}$ is the dominant component of the trace for all the value of $\gamma$ considered and in all the range of $D$, but for low value of the anisotropy parameter and high value of the DM interaction term it becomes comparable with the component of $H_{\gamma \gamma}$. 

The QFI of the anisotropy parameter, as $\gamma$ itself increases, becomes less and less significant. For negative values of $D$, $H_{\gamma \gamma}$ is less relevant than both $H_{JJ}$ and $H_{DD}$ and almost irrelevant independently on its value. For positive values of the DM interaction term, the QFI of $\gamma$ becomes more relevant than the element $H_{DD}$, independently on $\gamma$, and it is notable only when the parameter $D$ has an opposite sign with respect to the coupling constant $J$. Independently on $\gamma$, for low value of $D$, the $H_{DD}$ element of the QFIM is more relevant than the $\gamma$ component, while for high value of $D$ the behavior is reversed. The exact point in which the behavior is reversed depend on $\gamma$, and increases as the anisotropy increases.

\begin{figure}[h!]
    \centering
    \includegraphics[width=0.9\columnwidth]{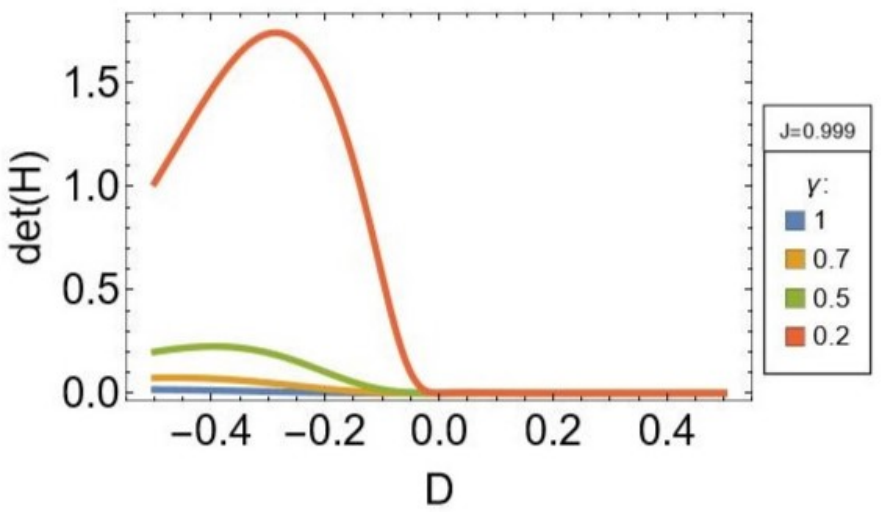}
    \caption{Determinant of the QFIM for a two-spin reduced density matrix 
    of an anisotropy spin chain with DM interaction as a function of $D$ 
    and different values of $\gamma$. The coupling is set to $J=0.999$.}
    \label{fig:fig:det_qfi_matrix_gamma_D}
\end{figure}

A global quantity, which summarize the above findings, may be obtained evaluating the determinant of the QFI matrix (see Fig. \ref{fig:fig:det_qfi_matrix_gamma_D}). For low value of the anisotropy parameter (i.e. $\gamma=0.2$) the determinant is greater than zero for negative $D$ and drops 
to zero for $D \geq 0$. This means that for $D<0$ all the parameters may be estimated separately, otherwise the QFI matrix is singular and the model is said to be \textit{sloppy}, because the state of the system is sensitive only to a combination of parameters \cite{gutenkunst2007universally} rather than on them separately. For higher 
values of the anisotropy $\gamma$, we have  $det(H) \simeq 0$ for the whole range of 
$D$ values. On the other hand, the elements of the Uhlmann matrix vanish  $U_{\mu \nu} = 0$. This means that in the joint estimation of any two parameters of the system, there is no intrinsic noise related to the non commutability of the symmetric logarithmic derivatives of $J$, $\gamma$ or $D$.

\subsection{Joint estimation at fixed anisotropy}
As a further example, we address the joint estimation 
of the three parameters $J$, $\gamma$ and $D$ assuming that the true 
value of the anisotropy is $\gamma=1$ (i.e., we focus to an Ising chain).
For concreteness' sake, we study  the behaviour of the QFIM for $J\in [-2,2]$  
and a discrete set of values of the DM coupling, $D=0.01, 0.1, 0.2, 0.3$.

\begin{figure}[h!]
    \centering
    \includegraphics[width=0.9\columnwidth]{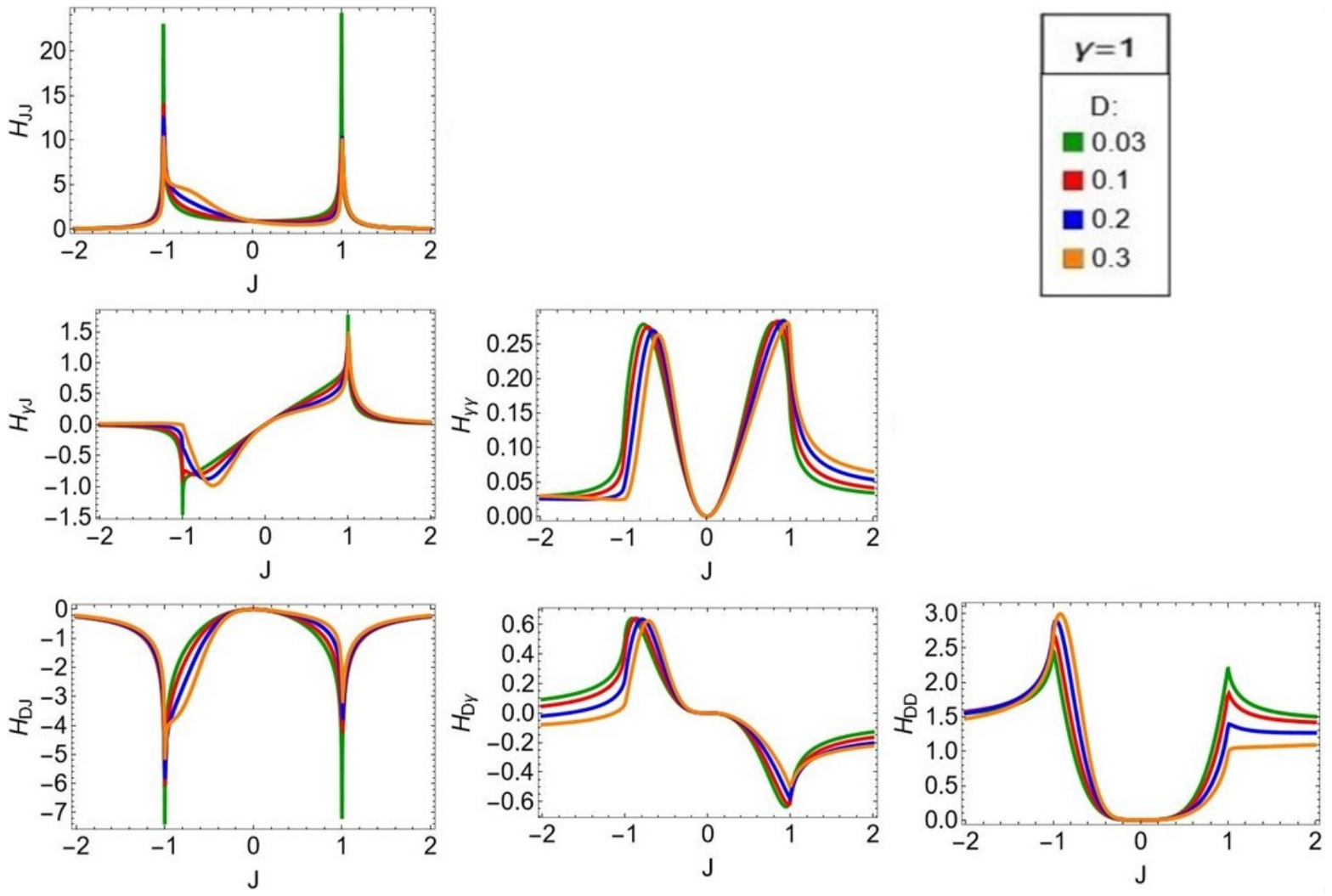}
    \caption{
    The matrix elements of the QFIM $H(J,\gamma,D)$ for a two-spin reduced density 
    matrix of an anisotropy spin chain as a function of $J$, and for different values of 
    $D$ (the anisotropy is set to $\gamma=1$). The array of plots corresponds to the position of the elements $H_{\mu \nu}$ in the QFIM, and the different curves in each plot denote the results for different values of the DM coupling $D$ (see the legend).}
    \label{fig:qfi_matrix_ising}
\end{figure}

As we can see from Fig. \ref{fig:qfi_matrix_ising}, the diagonal elements of the 
QFIM are larger in the regions around $J=\pm 1$. Notice however that the maximum 
of $H_{\gamma \gamma}$ is not exactly at $J=\pm1$, as for $H_{JJ}$ and $H_{DD}$, 
but at a value $J=\pm J^{*}$ with $\abs{J^{*}} < 1$. Looking at the non diagonal elements, we notice that they show some peak structure around $J = \pm 1$, that can be a maximum, as it happens for $H_{J\gamma}(J^{*}_{J\gamma})$ or $H_{\gamma D}(-J^{*})$ 
or a minimum, as for $H_{J\gamma}(-J^{*}_{J \gamma})$, $H_{\gamma D}(-J^{*}_{\gamma D})$ or $H_{JD}(\pm J^{*}_{JD})$. The main difference among the three diagonal elements 
is their magnitude, with $H_{DD}$ much larger than the other elements. As it happens 
in the single parameter case, the increase of $D$ modifies the height or the width
of the peaks, and this feature may be exploited to create a more robust or accurate measurement procedure.

\begin{figure}[h!]
    \centering
    \includegraphics[width=0.9\columnwidth]{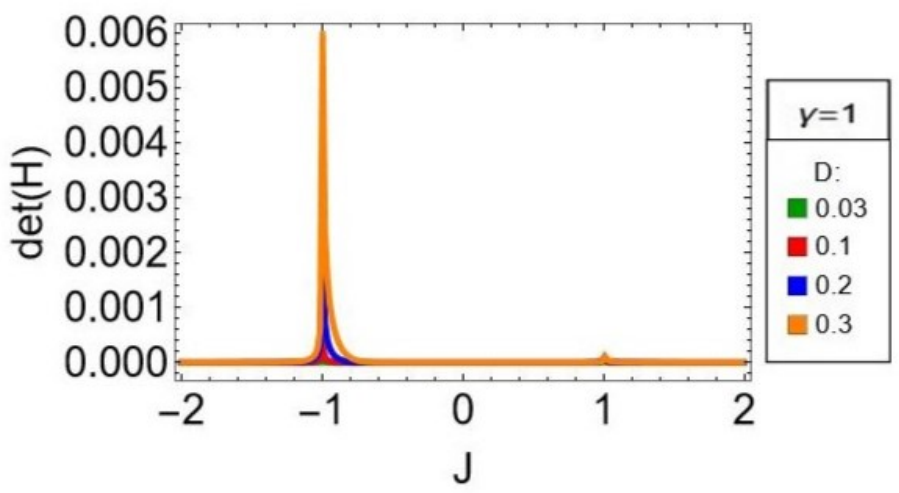}
    \caption{Determinant of the QFIM for a two-spin reduced density matrix 
    of an anisotropy spin chain with DM interaction as a function of $J$ 
    and different values of $D$. The anisotropy is set to $\gamma=1.0$.}
    \label{fig:det_qfi_matrix_ising}
\end{figure}

The determinant of the QFIM, as shown in Fig.\ref{fig:det_qfi_matrix_ising}, is very small in all the considered cases, thus confirming the \textit{sloppiness} of the statistical model. As previously reported for the joint estimation at fixed coupling, the elements of the Uhlmann matrix are again identically null, $U_{\mu \nu} = 0$. Then, also in this situation, the symmetric logarithmic derivatives of $J$, $\gamma$ or $D$ commute and this lead to an absence of intrinsic noise in the joint estimation at fixed anisotropy.

\section{Conclusions}\label{sec:Conclusions}
We have addressed the characterization of partially accessible 
anisotropic spin chains in the presence of anti-symmetric exchange, 
and explored the regimes where the Hamiltonian parameters of the chain 
may be estimated with precision approaching the ultimate limit imposed by 
quantum mechanics.  At variance with previous approaches, we have analyzed
the information that may be extracted by measuring only two neighbouring spins, which are 
used as quantum probes for the properties of the entire chain. 

Our results prove that measuring the total magnetization of the two-spin system may be indeed exploited to precisely estimate the parameters of the chain.
In particular, we have found that the ratio $S(J)=F(J)/H(J)$ between the magnetization FI and the corresponding QFI is close 
to unit for a large range of the coupling values. The presence of DM interaction improves the estimation of the coupling, since it leads to the presence of additional bumps and peaks in the FI and QFI, which may be exploited to increase the robustness of the overall estimation procedure. 

We have also addressed the multiparameter estimation problem, i.e. the joint estimation of the 
three parameters, and studied the dependence of the elements of the QFI matrix on the coupling and on the DM parameter. Our results show that the three parameters are compatible, i.e., the Uhlmann curvature vanishes and there is no additional noise of quantum origin. On the other hand, the model is sloppy, i.e., the state of the system is sensitive only to a combination of the parameters  rather than on them separately. 

Our results establish DM interaction as a resource for spin-chain metrology and pave the way to the development of scrambling procedures to remove the sloppiness of the model.

\begin{acknowledgments}
Work was done under the auspices of GNFM-INdAM. This work has been partially supported by MIUR through the project PRIN22-2022T25TR3-RISQUE, and by KU through the project C2PS-8474000137.
\end{acknowledgments}

\appendix

\bibliography{PDM.bib}

\begin{thebibliography}{60}%
\makeatletter
\providecommand \@ifxundefined [1]{%
 \@ifx{#1\undefined}
}%
\providecommand \@ifnum [1]{%
 \ifnum #1\expandafter \@firstoftwo
 \else \expandafter \@secondoftwo
 \fi
}%
\providecommand \@ifx [1]{%
 \ifx #1\expandafter \@firstoftwo
 \else \expandafter \@secondoftwo
 \fi
}%
\providecommand \natexlab [1]{#1}%
\providecommand \enquote  [1]{``#1''}%
\providecommand \bibnamefont  [1]{#1}%
\providecommand \bibfnamefont [1]{#1}%
\providecommand \citenamefont [1]{#1}%
\providecommand \href@noop [0]{\@secondoftwo}%
\providecommand \href [0]{\begingroup \@sanitize@url \@href}%
\providecommand \@href[1]{\@@startlink{#1}\@@href}%
\providecommand \@@href[1]{\endgroup#1\@@endlink}%
\providecommand \@sanitize@url [0]{\catcode `\\12\catcode `\$12\catcode
  `\&12\catcode `\#12\catcode `\^12\catcode `\_12\catcode `\%12\relax}%
\providecommand \@@startlink[1]{}%
\providecommand \@@endlink[0]{}%
\providecommand \url  [0]{\begingroup\@sanitize@url \@url }%
\providecommand \@url [1]{\endgroup\@href {#1}{\urlprefix }}%
\providecommand \urlprefix  [0]{URL }%
\providecommand \Eprint [0]{\href }%
\providecommand \doibase [0]{https://doi.org/}%
\providecommand \selectlanguage [0]{\@gobble}%
\providecommand \bibinfo  [0]{\@secondoftwo}%
\providecommand \bibfield  [0]{\@secondoftwo}%
\providecommand \translation [1]{[#1]}%
\providecommand \BibitemOpen [0]{}%
\providecommand \bibitemStop [0]{}%
\providecommand \bibitemNoStop [0]{.\EOS\space}%
\providecommand \EOS [0]{\spacefactor3000\relax}%
\providecommand \BibitemShut  [1]{\csname bibitem#1\endcsname}%
\let\auto@bib@innerbib\@empty
\bibitem [{\citenamefont {Paris}\ and\ \citenamefont
  {Rehacek}(2004)}]{paris2004quantum}%
  \BibitemOpen
  \bibfield  {author} {\bibinfo {author} {\bibfnamefont {M.~G.~A.}\
  \bibnamefont {Paris}}\ and\ \bibinfo {author} {\bibfnamefont
  {J.}~\bibnamefont {Rehacek}},\ }\href@noop {} {\emph {\bibinfo {title}
  {Quantum state estimation}}},\ Vol.\ \bibinfo {volume} {649}\ (\bibinfo
  {publisher} {Springer Science \& Business Media},\ \bibinfo {year}
  {2004})\BibitemShut {NoStop}%
\bibitem [{\citenamefont {Petz}\ and\ \citenamefont
  {Ghinea}(2011)}]{petz2011introduction}%
  \BibitemOpen
  \bibfield  {author} {\bibinfo {author} {\bibfnamefont {D.}~\bibnamefont
  {Petz}}\ and\ \bibinfo {author} {\bibfnamefont {C.}~\bibnamefont {Ghinea}},\
  }in\ \href@noop {} {\emph {\bibinfo {booktitle} {Quantum probability and
  related topics}}}\ (\bibinfo  {publisher} {World Scientific},\ \bibinfo
  {year} {2011})\ pp.\ \bibinfo {pages} {261--281}\BibitemShut {NoStop}%
\bibitem [{\citenamefont {Invernizzi}\ \emph {et~al.}(2008)\citenamefont
  {Invernizzi}, \citenamefont {Korbman}, \citenamefont {Venuti},\ and\
  \citenamefont {Paris}}]{invernizzi2008optimal}%
  \BibitemOpen
  \bibfield  {author} {\bibinfo {author} {\bibfnamefont {C.}~\bibnamefont
  {Invernizzi}}, \bibinfo {author} {\bibfnamefont {M.}~\bibnamefont {Korbman}},
  \bibinfo {author} {\bibfnamefont {L.~C.}\ \bibnamefont {Venuti}},\ and\
  \bibinfo {author} {\bibfnamefont {M.~G.}\ \bibnamefont {Paris}},\ }\href@noop
  {} {\bibfield  {journal} {\bibinfo  {journal} {Physical Review A}\ }\textbf
  {\bibinfo {volume} {78}},\ \bibinfo {pages} {042106} (\bibinfo {year}
  {2008})}\BibitemShut {NoStop}%
\bibitem [{\citenamefont {Zanardi}\ \emph {et~al.}(2008)\citenamefont
  {Zanardi}, \citenamefont {Paris},\ and\ \citenamefont
  {Venuti}}]{zanardi2008quantum}%
  \BibitemOpen
  \bibfield  {author} {\bibinfo {author} {\bibfnamefont {P.}~\bibnamefont
  {Zanardi}}, \bibinfo {author} {\bibfnamefont {M.~G.}\ \bibnamefont {Paris}},\
  and\ \bibinfo {author} {\bibfnamefont {L.~C.}\ \bibnamefont {Venuti}},\
  }\href@noop {} {\bibfield  {journal} {\bibinfo  {journal} {Physical Review
  A}\ }\textbf {\bibinfo {volume} {78}},\ \bibinfo {pages} {042105} (\bibinfo
  {year} {2008})}\BibitemShut {NoStop}%
\bibitem [{\citenamefont {Chu}\ \emph {et~al.}(2021)\citenamefont {Chu},
  \citenamefont {Zhang}, \citenamefont {Yu},\ and\ \citenamefont
  {Cai}}]{chu2021dynamic}%
  \BibitemOpen
  \bibfield  {author} {\bibinfo {author} {\bibfnamefont {Y.}~\bibnamefont
  {Chu}}, \bibinfo {author} {\bibfnamefont {S.}~\bibnamefont {Zhang}}, \bibinfo
  {author} {\bibfnamefont {B.}~\bibnamefont {Yu}},\ and\ \bibinfo {author}
  {\bibfnamefont {J.}~\bibnamefont {Cai}},\ }\href@noop {} {\bibfield
  {journal} {\bibinfo  {journal} {Physical Review Letters}\ }\textbf {\bibinfo
  {volume} {126}},\ \bibinfo {pages} {010502} (\bibinfo {year}
  {2021})}\BibitemShut {NoStop}%
\bibitem [{\citenamefont {Cepas}\ \emph {et~al.}(2008)\citenamefont {Cepas},
  \citenamefont {Fong}, \citenamefont {Leung},\ and\ \citenamefont
  {Lhuillier}}]{cepas2008quantum}%
  \BibitemOpen
  \bibfield  {author} {\bibinfo {author} {\bibfnamefont {O.}~\bibnamefont
  {Cepas}}, \bibinfo {author} {\bibfnamefont {C.}~\bibnamefont {Fong}},
  \bibinfo {author} {\bibfnamefont {P.~W.}\ \bibnamefont {Leung}},\ and\
  \bibinfo {author} {\bibfnamefont {C.}~\bibnamefont {Lhuillier}},\ }\href@noop
  {} {\bibfield  {journal} {\bibinfo  {journal} {Physical Review B}\ }\textbf
  {\bibinfo {volume} {78}},\ \bibinfo {pages} {140405} (\bibinfo {year}
  {2008})}\BibitemShut {NoStop}%
\bibitem [{\citenamefont {Jafari}\ \emph {et~al.}(2008)\citenamefont {Jafari},
  \citenamefont {Kargarian}, \citenamefont {Langari},\ and\ \citenamefont
  {Siahatgar}}]{jafari2008phase}%
  \BibitemOpen
  \bibfield  {author} {\bibinfo {author} {\bibfnamefont {R.}~\bibnamefont
  {Jafari}}, \bibinfo {author} {\bibfnamefont {M.}~\bibnamefont {Kargarian}},
  \bibinfo {author} {\bibfnamefont {A.}~\bibnamefont {Langari}},\ and\ \bibinfo
  {author} {\bibfnamefont {M.}~\bibnamefont {Siahatgar}},\ }\href@noop {}
  {\bibfield  {journal} {\bibinfo  {journal} {Physical Review B}\ }\textbf
  {\bibinfo {volume} {78}},\ \bibinfo {pages} {214414} (\bibinfo {year}
  {2008})}\BibitemShut {NoStop}%
\bibitem [{\citenamefont {Jin}\ and\ \citenamefont
  {Starykh}(2017)}]{jin2017phase}%
  \BibitemOpen
  \bibfield  {author} {\bibinfo {author} {\bibfnamefont {W.}~\bibnamefont
  {Jin}}\ and\ \bibinfo {author} {\bibfnamefont {O.~A.}\ \bibnamefont
  {Starykh}},\ }\href@noop {} {\bibfield  {journal} {\bibinfo  {journal}
  {Physical Review B}\ }\textbf {\bibinfo {volume} {95}},\ \bibinfo {pages}
  {214404} (\bibinfo {year} {2017})}\BibitemShut {NoStop}%
\bibitem [{\citenamefont {Sergienko}\ and\ \citenamefont
  {Dagotto}(2006)}]{sergienko2006role}%
  \BibitemOpen
  \bibfield  {author} {\bibinfo {author} {\bibfnamefont {I.~A.}\ \bibnamefont
  {Sergienko}}\ and\ \bibinfo {author} {\bibfnamefont {E.}~\bibnamefont
  {Dagotto}},\ }\href@noop {} {\bibfield  {journal} {\bibinfo  {journal}
  {Physical Review B}\ }\textbf {\bibinfo {volume} {73}},\ \bibinfo {pages}
  {094434} (\bibinfo {year} {2006})}\BibitemShut {NoStop}%
\bibitem [{\citenamefont {H{\"a}lg}\ \emph {et~al.}(2014)\citenamefont
  {H{\"a}lg}, \citenamefont {Lorenz}, \citenamefont {Povarov}, \citenamefont
  {M{\aa}nsson}, \citenamefont {Skourski},\ and\ \citenamefont
  {Zheludev}}]{halg2014quantum}%
  \BibitemOpen
  \bibfield  {author} {\bibinfo {author} {\bibfnamefont {M.}~\bibnamefont
  {H{\"a}lg}}, \bibinfo {author} {\bibfnamefont {W.~E.}\ \bibnamefont
  {Lorenz}}, \bibinfo {author} {\bibfnamefont {K.~Y.}\ \bibnamefont {Povarov}},
  \bibinfo {author} {\bibfnamefont {M.}~\bibnamefont {M{\aa}nsson}}, \bibinfo
  {author} {\bibfnamefont {Y.}~\bibnamefont {Skourski}},\ and\ \bibinfo
  {author} {\bibfnamefont {A.}~\bibnamefont {Zheludev}},\ }\href@noop {}
  {\bibfield  {journal} {\bibinfo  {journal} {Physical Review B}\ }\textbf
  {\bibinfo {volume} {90}},\ \bibinfo {pages} {174413} (\bibinfo {year}
  {2014})}\BibitemShut {NoStop}%
\bibitem [{\citenamefont {Yang}\ \emph {et~al.}(2015)\citenamefont {Yang},
  \citenamefont {Thiaville}, \citenamefont {Rohart}, \citenamefont {Fert},\
  and\ \citenamefont {Chshiev}}]{yang2015anatomy}%
  \BibitemOpen
  \bibfield  {author} {\bibinfo {author} {\bibfnamefont {H.}~\bibnamefont
  {Yang}}, \bibinfo {author} {\bibfnamefont {A.}~\bibnamefont {Thiaville}},
  \bibinfo {author} {\bibfnamefont {S.}~\bibnamefont {Rohart}}, \bibinfo
  {author} {\bibfnamefont {A.}~\bibnamefont {Fert}},\ and\ \bibinfo {author}
  {\bibfnamefont {M.}~\bibnamefont {Chshiev}},\ }\href@noop {} {\bibfield
  {journal} {\bibinfo  {journal} {Physical review letters}\ }\textbf {\bibinfo
  {volume} {115}},\ \bibinfo {pages} {267210} (\bibinfo {year}
  {2015})}\BibitemShut {NoStop}%
\bibitem [{\citenamefont {Derzhko}\ \emph {et~al.}(2006)\citenamefont
  {Derzhko}, \citenamefont {Verkholyak}, \citenamefont {Krokhmalskii},\ and\
  \citenamefont {B{\"u}ttner}}]{derzhko2006dynamic}%
  \BibitemOpen
  \bibfield  {author} {\bibinfo {author} {\bibfnamefont {O.}~\bibnamefont
  {Derzhko}}, \bibinfo {author} {\bibfnamefont {T.}~\bibnamefont {Verkholyak}},
  \bibinfo {author} {\bibfnamefont {T.}~\bibnamefont {Krokhmalskii}},\ and\
  \bibinfo {author} {\bibfnamefont {H.}~\bibnamefont {B{\"u}ttner}},\
  }\href@noop {} {\bibfield  {journal} {\bibinfo  {journal} {Physical Review
  B}\ }\textbf {\bibinfo {volume} {73}},\ \bibinfo {pages} {214407} (\bibinfo
  {year} {2006})}\BibitemShut {NoStop}%
\bibitem [{\citenamefont {Gangadharaiah}\ \emph {et~al.}(2008)\citenamefont
  {Gangadharaiah}, \citenamefont {Sun},\ and\ \citenamefont
  {Starykh}}]{gangadharaiah2008spin}%
  \BibitemOpen
  \bibfield  {author} {\bibinfo {author} {\bibfnamefont {S.}~\bibnamefont
  {Gangadharaiah}}, \bibinfo {author} {\bibfnamefont {J.}~\bibnamefont {Sun}},\
  and\ \bibinfo {author} {\bibfnamefont {O.~A.}\ \bibnamefont {Starykh}},\
  }\href@noop {} {\bibfield  {journal} {\bibinfo  {journal} {Physical Review
  B}\ }\textbf {\bibinfo {volume} {78}},\ \bibinfo {pages} {054436} (\bibinfo
  {year} {2008})}\BibitemShut {NoStop}%
\bibitem [{\citenamefont {Chan}\ \emph {et~al.}(2017)\citenamefont {Chan},
  \citenamefont {Jin}, \citenamefont {Jiang},\ and\ \citenamefont
  {Starykh}}]{chan2017ising}%
  \BibitemOpen
  \bibfield  {author} {\bibinfo {author} {\bibfnamefont {Y.-H.}\ \bibnamefont
  {Chan}}, \bibinfo {author} {\bibfnamefont {W.}~\bibnamefont {Jin}}, \bibinfo
  {author} {\bibfnamefont {H.-C.}\ \bibnamefont {Jiang}},\ and\ \bibinfo
  {author} {\bibfnamefont {O.~A.}\ \bibnamefont {Starykh}},\ }\href@noop {}
  {\bibfield  {journal} {\bibinfo  {journal} {Physical Review B}\ }\textbf
  {\bibinfo {volume} {96}},\ \bibinfo {pages} {214441} (\bibinfo {year}
  {2017})}\BibitemShut {NoStop}%
\bibitem [{\citenamefont {Pylypovskyi}\ \emph {et~al.}(2021)\citenamefont
  {Pylypovskyi}, \citenamefont {Borysenko}, \citenamefont {Fassbender},
  \citenamefont {Sheka},\ and\ \citenamefont
  {Makarov}}]{pylypovskyi2021curvature}%
  \BibitemOpen
  \bibfield  {author} {\bibinfo {author} {\bibfnamefont {O.~V.}\ \bibnamefont
  {Pylypovskyi}}, \bibinfo {author} {\bibfnamefont {Y.~A.}\ \bibnamefont
  {Borysenko}}, \bibinfo {author} {\bibfnamefont {J.}~\bibnamefont
  {Fassbender}}, \bibinfo {author} {\bibfnamefont {D.~D.}\ \bibnamefont
  {Sheka}},\ and\ \bibinfo {author} {\bibfnamefont {D.}~\bibnamefont
  {Makarov}},\ }\href@noop {} {\bibfield  {journal} {\bibinfo  {journal}
  {Applied Physics Letters}\ }\textbf {\bibinfo {volume} {118}} (\bibinfo
  {year} {2021})}\BibitemShut {NoStop}%
\bibitem [{\citenamefont {Fumani}\ \emph {et~al.}(2021)\citenamefont {Fumani},
  \citenamefont {Beradze}, \citenamefont {Nemati}, \citenamefont {Mahdavifar},\
  and\ \citenamefont {Japaridze}}]{fumani2021quantum}%
  \BibitemOpen
  \bibfield  {author} {\bibinfo {author} {\bibfnamefont {F.~K.}\ \bibnamefont
  {Fumani}}, \bibinfo {author} {\bibfnamefont {B.}~\bibnamefont {Beradze}},
  \bibinfo {author} {\bibfnamefont {S.}~\bibnamefont {Nemati}}, \bibinfo
  {author} {\bibfnamefont {S.}~\bibnamefont {Mahdavifar}},\ and\ \bibinfo
  {author} {\bibfnamefont {G.}~\bibnamefont {Japaridze}},\ }\href@noop {}
  {\bibfield  {journal} {\bibinfo  {journal} {Journal of Magnetism and Magnetic
  Materials}\ }\textbf {\bibinfo {volume} {518}},\ \bibinfo {pages} {167411}
  (\bibinfo {year} {2021})}\BibitemShut {NoStop}%
\bibitem [{\citenamefont {Pham}\ \emph {et~al.}(2021)\citenamefont {Pham},
  \citenamefont {Ngo}, \citenamefont {Le}, \citenamefont {Hoang}, \citenamefont
  {Phan},\ and\ \citenamefont {Nguyen}}]{pham2021effect}%
  \BibitemOpen
  \bibfield  {author} {\bibinfo {author} {\bibfnamefont {H.~T.}\ \bibnamefont
  {Pham}}, \bibinfo {author} {\bibfnamefont {T.~T.}\ \bibnamefont {Ngo}},
  \bibinfo {author} {\bibfnamefont {T.~T.}\ \bibnamefont {Le}}, \bibinfo
  {author} {\bibfnamefont {D.~L.}\ \bibnamefont {Hoang}}, \bibinfo {author}
  {\bibfnamefont {T.~N.}\ \bibnamefont {Phan}},\ and\ \bibinfo {author}
  {\bibfnamefont {H.~C.}\ \bibnamefont {Nguyen}},\ }\href@noop {} {\bibfield
  {journal} {\bibinfo  {journal} {Hue University Journal of Science: Natural
  Science}\ }\textbf {\bibinfo {volume} {130}},\ \bibinfo {pages} {31}
  (\bibinfo {year} {2021})}\BibitemShut {NoStop}%
\bibitem [{\citenamefont {Di}\ \emph {et~al.}(2015)\citenamefont {Di},
  \citenamefont {Zhang}, \citenamefont {Lim}, \citenamefont {Ng}, \citenamefont
  {Kuok}, \citenamefont {Yu}, \citenamefont {Yoon}, \citenamefont {Qiu},\ and\
  \citenamefont {Yang}}]{di2015direct}%
  \BibitemOpen
  \bibfield  {author} {\bibinfo {author} {\bibfnamefont {K.}~\bibnamefont
  {Di}}, \bibinfo {author} {\bibfnamefont {V.~L.}\ \bibnamefont {Zhang}},
  \bibinfo {author} {\bibfnamefont {H.~S.}\ \bibnamefont {Lim}}, \bibinfo
  {author} {\bibfnamefont {S.~C.}\ \bibnamefont {Ng}}, \bibinfo {author}
  {\bibfnamefont {M.~H.}\ \bibnamefont {Kuok}}, \bibinfo {author}
  {\bibfnamefont {J.}~\bibnamefont {Yu}}, \bibinfo {author} {\bibfnamefont
  {J.}~\bibnamefont {Yoon}}, \bibinfo {author} {\bibfnamefont {X.}~\bibnamefont
  {Qiu}},\ and\ \bibinfo {author} {\bibfnamefont {H.}~\bibnamefont {Yang}},\
  }\href@noop {} {\bibfield  {journal} {\bibinfo  {journal} {Physical review
  letters}\ }\textbf {\bibinfo {volume} {114}},\ \bibinfo {pages} {047201}
  (\bibinfo {year} {2015})}\BibitemShut {NoStop}%
\bibitem [{\citenamefont {Dmitrienko}\ \emph {et~al.}(2014)\citenamefont
  {Dmitrienko}, \citenamefont {Ovchinnikova}, \citenamefont {Collins},
  \citenamefont {Nisbet}, \citenamefont {Beutier}, \citenamefont {Kvashnin},
  \citenamefont {Mazurenko}, \citenamefont {Lichtenstein},\ and\ \citenamefont
  {Katsnelson}}]{dmitrienko2014measuring}%
  \BibitemOpen
  \bibfield  {author} {\bibinfo {author} {\bibfnamefont {V.}~\bibnamefont
  {Dmitrienko}}, \bibinfo {author} {\bibfnamefont {E.}~\bibnamefont
  {Ovchinnikova}}, \bibinfo {author} {\bibfnamefont {S.}~\bibnamefont
  {Collins}}, \bibinfo {author} {\bibfnamefont {G.}~\bibnamefont {Nisbet}},
  \bibinfo {author} {\bibfnamefont {G.}~\bibnamefont {Beutier}}, \bibinfo
  {author} {\bibfnamefont {Y.}~\bibnamefont {Kvashnin}}, \bibinfo {author}
  {\bibfnamefont {V.}~\bibnamefont {Mazurenko}}, \bibinfo {author}
  {\bibfnamefont {A.}~\bibnamefont {Lichtenstein}},\ and\ \bibinfo {author}
  {\bibfnamefont {M.}~\bibnamefont {Katsnelson}},\ }\href@noop {} {\bibfield
  {journal} {\bibinfo  {journal} {Nature Physics}\ }\textbf {\bibinfo {volume}
  {10}},\ \bibinfo {pages} {202} (\bibinfo {year} {2014})}\BibitemShut
  {NoStop}%
\bibitem [{\citenamefont {Yang}\ \emph {et~al.}(2023)\citenamefont {Yang},
  \citenamefont {Liang},\ and\ \citenamefont {Cui}}]{yang2023first}%
  \BibitemOpen
  \bibfield  {author} {\bibinfo {author} {\bibfnamefont {H.}~\bibnamefont
  {Yang}}, \bibinfo {author} {\bibfnamefont {J.}~\bibnamefont {Liang}},\ and\
  \bibinfo {author} {\bibfnamefont {Q.}~\bibnamefont {Cui}},\ }\href@noop {}
  {\bibfield  {journal} {\bibinfo  {journal} {Nature Reviews Physics}\ }\textbf
  {\bibinfo {volume} {5}},\ \bibinfo {pages} {43} (\bibinfo {year}
  {2023})}\BibitemShut {NoStop}%
\bibitem [{\citenamefont {Yi}\ \emph {et~al.}(2019)\citenamefont {Yi},
  \citenamefont {You}, \citenamefont {Wu},\ and\ \citenamefont
  {Ole{\'s}}}]{yi2019criticality}%
  \BibitemOpen
  \bibfield  {author} {\bibinfo {author} {\bibfnamefont {T.-C.}\ \bibnamefont
  {Yi}}, \bibinfo {author} {\bibfnamefont {W.-L.}\ \bibnamefont {You}},
  \bibinfo {author} {\bibfnamefont {N.}~\bibnamefont {Wu}},\ and\ \bibinfo
  {author} {\bibfnamefont {A.~M.}\ \bibnamefont {Ole{\'s}}},\ }\href@noop {}
  {\bibfield  {journal} {\bibinfo  {journal} {Physical Review B}\ }\textbf
  {\bibinfo {volume} {100}},\ \bibinfo {pages} {024423} (\bibinfo {year}
  {2019})}\BibitemShut {NoStop}%
\bibitem [{\citenamefont {Ait~Chlih}\ \emph {et~al.}(2021)\citenamefont
  {Ait~Chlih}, \citenamefont {Habiballah},\ and\ \citenamefont
  {Nassik}}]{ait2021dynamics}%
  \BibitemOpen
  \bibfield  {author} {\bibinfo {author} {\bibfnamefont {A.}~\bibnamefont
  {Ait~Chlih}}, \bibinfo {author} {\bibfnamefont {N.}~\bibnamefont
  {Habiballah}},\ and\ \bibinfo {author} {\bibfnamefont {M.}~\bibnamefont
  {Nassik}},\ }\href@noop {} {\bibfield  {journal} {\bibinfo  {journal}
  {Quantum Information Processing}\ }\textbf {\bibinfo {volume} {20}},\
  \bibinfo {pages} {1} (\bibinfo {year} {2021})}\BibitemShut {NoStop}%
\bibitem [{\citenamefont {Liang}\ \emph {et~al.}(2022)\citenamefont {Liang},
  \citenamefont {Chshiev}, \citenamefont {Fert},\ and\ \citenamefont
  {Yang}}]{liang2022gradient}%
  \BibitemOpen
  \bibfield  {author} {\bibinfo {author} {\bibfnamefont {J.}~\bibnamefont
  {Liang}}, \bibinfo {author} {\bibfnamefont {M.}~\bibnamefont {Chshiev}},
  \bibinfo {author} {\bibfnamefont {A.}~\bibnamefont {Fert}},\ and\ \bibinfo
  {author} {\bibfnamefont {H.}~\bibnamefont {Yang}},\ }\href@noop {} {\bibfield
   {journal} {\bibinfo  {journal} {Nano Letters}\ }\textbf {\bibinfo {volume}
  {22}},\ \bibinfo {pages} {10128} (\bibinfo {year} {2022})}\BibitemShut
  {NoStop}%
\bibitem [{\citenamefont {Kuepferling}\ \emph {et~al.}(2023)\citenamefont
  {Kuepferling}, \citenamefont {Casiraghi}, \citenamefont {Soares},
  \citenamefont {Durin}, \citenamefont {Garcia-Sanchez}, \citenamefont {Chen},
  \citenamefont {Back}, \citenamefont {Marrows}, \citenamefont {Tacchi},\ and\
  \citenamefont {Carlotti}}]{kuepferling2023measuring}%
  \BibitemOpen
  \bibfield  {author} {\bibinfo {author} {\bibfnamefont {M.}~\bibnamefont
  {Kuepferling}}, \bibinfo {author} {\bibfnamefont {A.}~\bibnamefont
  {Casiraghi}}, \bibinfo {author} {\bibfnamefont {G.}~\bibnamefont {Soares}},
  \bibinfo {author} {\bibfnamefont {G.}~\bibnamefont {Durin}}, \bibinfo
  {author} {\bibfnamefont {F.}~\bibnamefont {Garcia-Sanchez}}, \bibinfo
  {author} {\bibfnamefont {L.}~\bibnamefont {Chen}}, \bibinfo {author}
  {\bibfnamefont {C.~H.}\ \bibnamefont {Back}}, \bibinfo {author}
  {\bibfnamefont {C.~H.}\ \bibnamefont {Marrows}}, \bibinfo {author}
  {\bibfnamefont {S.}~\bibnamefont {Tacchi}},\ and\ \bibinfo {author}
  {\bibfnamefont {G.}~\bibnamefont {Carlotti}},\ }\href@noop {} {\bibfield
  {journal} {\bibinfo  {journal} {Reviews of Modern Physics}\ }\textbf
  {\bibinfo {volume} {95}},\ \bibinfo {pages} {015003} (\bibinfo {year}
  {2023})}\BibitemShut {NoStop}%
\bibitem [{\citenamefont {Gusev}\ \emph {et~al.}(2020)\citenamefont {Gusev},
  \citenamefont {Sadovnikov}, \citenamefont {Nikitov}, \citenamefont
  {Sapozhnikov},\ and\ \citenamefont {Udalov}}]{gusev2020manipulation}%
  \BibitemOpen
  \bibfield  {author} {\bibinfo {author} {\bibfnamefont {N.}~\bibnamefont
  {Gusev}}, \bibinfo {author} {\bibfnamefont {A.}~\bibnamefont {Sadovnikov}},
  \bibinfo {author} {\bibfnamefont {S.}~\bibnamefont {Nikitov}}, \bibinfo
  {author} {\bibfnamefont {M.}~\bibnamefont {Sapozhnikov}},\ and\ \bibinfo
  {author} {\bibfnamefont {O.}~\bibnamefont {Udalov}},\ }\href@noop {}
  {\bibfield  {journal} {\bibinfo  {journal} {Physical review letters}\
  }\textbf {\bibinfo {volume} {124}},\ \bibinfo {pages} {157202} (\bibinfo
  {year} {2020})}\BibitemShut {NoStop}%
\bibitem [{\citenamefont {Zhang}\ \emph {et~al.}(2022)\citenamefont {Zhang},
  \citenamefont {Liang}, \citenamefont {Bi}, \citenamefont {Zhao},
  \citenamefont {Bai}, \citenamefont {Cui}, \citenamefont {Zhou}, \citenamefont
  {Bai}, \citenamefont {Feng}, \citenamefont {Song} \emph
  {et~al.}}]{zhang2022quantifying}%
  \BibitemOpen
  \bibfield  {author} {\bibinfo {author} {\bibfnamefont {Q.}~\bibnamefont
  {Zhang}}, \bibinfo {author} {\bibfnamefont {J.}~\bibnamefont {Liang}},
  \bibinfo {author} {\bibfnamefont {K.}~\bibnamefont {Bi}}, \bibinfo {author}
  {\bibfnamefont {L.}~\bibnamefont {Zhao}}, \bibinfo {author} {\bibfnamefont
  {H.}~\bibnamefont {Bai}}, \bibinfo {author} {\bibfnamefont {Q.}~\bibnamefont
  {Cui}}, \bibinfo {author} {\bibfnamefont {H.-A.}\ \bibnamefont {Zhou}},
  \bibinfo {author} {\bibfnamefont {H.}~\bibnamefont {Bai}}, \bibinfo {author}
  {\bibfnamefont {H.}~\bibnamefont {Feng}}, \bibinfo {author} {\bibfnamefont
  {W.}~\bibnamefont {Song}}, \emph {et~al.},\ }\href@noop {} {\bibfield
  {journal} {\bibinfo  {journal} {Physical Review Letters}\ }\textbf {\bibinfo
  {volume} {128}},\ \bibinfo {pages} {167202} (\bibinfo {year}
  {2022})}\BibitemShut {NoStop}%
\bibitem [{\citenamefont {Wu}\ and\ \citenamefont
  {Lidar}(2002)}]{wu2002universal}%
  \BibitemOpen
  \bibfield  {author} {\bibinfo {author} {\bibfnamefont {L.-A.}\ \bibnamefont
  {Wu}}\ and\ \bibinfo {author} {\bibfnamefont {D.~A.}\ \bibnamefont {Lidar}},\
  }\href@noop {} {\bibfield  {journal} {\bibinfo  {journal} {Physical Review
  A}\ }\textbf {\bibinfo {volume} {66}},\ \bibinfo {pages} {062314} (\bibinfo
  {year} {2002})}\BibitemShut {NoStop}%
\bibitem [{\citenamefont {Yang}\ \emph {et~al.}(2019)\citenamefont {Yang},
  \citenamefont {Sun}, \citenamefont {Shi}, \citenamefont {Ming}, \citenamefont
  {Wang},\ and\ \citenamefont {Ye}}]{yang2019dynamical}%
  \BibitemOpen
  \bibfield  {author} {\bibinfo {author} {\bibfnamefont {Y.-Y.}\ \bibnamefont
  {Yang}}, \bibinfo {author} {\bibfnamefont {W.-Y.}\ \bibnamefont {Sun}},
  \bibinfo {author} {\bibfnamefont {W.-N.}\ \bibnamefont {Shi}}, \bibinfo
  {author} {\bibfnamefont {F.}~\bibnamefont {Ming}}, \bibinfo {author}
  {\bibfnamefont {D.}~\bibnamefont {Wang}},\ and\ \bibinfo {author}
  {\bibfnamefont {L.}~\bibnamefont {Ye}},\ }\href@noop {} {\bibfield  {journal}
  {\bibinfo  {journal} {Frontiers of Physics}\ }\textbf {\bibinfo {volume}
  {14}},\ \bibinfo {pages} {1} (\bibinfo {year} {2019})}\BibitemShut {NoStop}%
\bibitem [{\citenamefont {Maruyama}\ \emph {et~al.}(2007)\citenamefont
  {Maruyama}, \citenamefont {Iitaka},\ and\ \citenamefont
  {Nori}}]{maruyama2007enhancement}%
  \BibitemOpen
  \bibfield  {author} {\bibinfo {author} {\bibfnamefont {K.}~\bibnamefont
  {Maruyama}}, \bibinfo {author} {\bibfnamefont {T.}~\bibnamefont {Iitaka}},\
  and\ \bibinfo {author} {\bibfnamefont {F.}~\bibnamefont {Nori}},\ }\href@noop
  {} {\bibfield  {journal} {\bibinfo  {journal} {Physical Review A}\ }\textbf
  {\bibinfo {volume} {75}},\ \bibinfo {pages} {012325} (\bibinfo {year}
  {2007})}\BibitemShut {NoStop}%
\bibitem [{\citenamefont {Zhang}(2007)}]{zhang2007thermal}%
  \BibitemOpen
  \bibfield  {author} {\bibinfo {author} {\bibfnamefont {G.-F.}\ \bibnamefont
  {Zhang}},\ }\href@noop {} {\bibfield  {journal} {\bibinfo  {journal}
  {Physical Review A}\ }\textbf {\bibinfo {volume} {75}},\ \bibinfo {pages}
  {034304} (\bibinfo {year} {2007})}\BibitemShut {NoStop}%
\bibitem [{\citenamefont {Chuan-Jia}\ \emph {et~al.}(2008)\citenamefont
  {Chuan-Jia}, \citenamefont {Wei-Wen}, \citenamefont {Tang-Kun}, \citenamefont
  {Yan-Xia},\ and\ \citenamefont {Hong}}]{chuan2008entanglement}%
  \BibitemOpen
  \bibfield  {author} {\bibinfo {author} {\bibfnamefont {S.}~\bibnamefont
  {Chuan-Jia}}, \bibinfo {author} {\bibfnamefont {C.}~\bibnamefont {Wei-Wen}},
  \bibinfo {author} {\bibfnamefont {L.}~\bibnamefont {Tang-Kun}}, \bibinfo
  {author} {\bibfnamefont {H.}~\bibnamefont {Yan-Xia}},\ and\ \bibinfo {author}
  {\bibfnamefont {L.}~\bibnamefont {Hong}},\ }\href@noop {} {\bibfield
  {journal} {\bibinfo  {journal} {Chinese Physics Letters}\ }\textbf {\bibinfo
  {volume} {25}},\ \bibinfo {pages} {817} (\bibinfo {year} {2008})}\BibitemShut
  {NoStop}%
\bibitem [{\citenamefont {Kargarian}\ \emph {et~al.}(2009)\citenamefont
  {Kargarian}, \citenamefont {Jafari},\ and\ \citenamefont
  {Langari}}]{kargarian2009dzyaloshinskii}%
  \BibitemOpen
  \bibfield  {author} {\bibinfo {author} {\bibfnamefont {M.}~\bibnamefont
  {Kargarian}}, \bibinfo {author} {\bibfnamefont {R.}~\bibnamefont {Jafari}},\
  and\ \bibinfo {author} {\bibfnamefont {A.}~\bibnamefont {Langari}},\
  }\href@noop {} {\bibfield  {journal} {\bibinfo  {journal} {Physical Review
  A}\ }\textbf {\bibinfo {volume} {79}},\ \bibinfo {pages} {042319} (\bibinfo
  {year} {2009})}\BibitemShut {NoStop}%
\bibitem [{\citenamefont {Park}(2019)}]{park2019thermal}%
  \BibitemOpen
  \bibfield  {author} {\bibinfo {author} {\bibfnamefont {D.}~\bibnamefont
  {Park}},\ }\href@noop {} {\bibfield  {journal} {\bibinfo  {journal} {Quantum
  Information Processing}\ }\textbf {\bibinfo {volume} {18}},\ \bibinfo {pages}
  {1} (\bibinfo {year} {2019})}\BibitemShut {NoStop}%
\bibitem [{\citenamefont {Liu}\ \emph {et~al.}(2011)\citenamefont {Liu},
  \citenamefont {Shao}, \citenamefont {Li}, \citenamefont {Zou},\ and\
  \citenamefont {Wu}}]{liu2011quantum}%
  \BibitemOpen
  \bibfield  {author} {\bibinfo {author} {\bibfnamefont {B.-Q.}\ \bibnamefont
  {Liu}}, \bibinfo {author} {\bibfnamefont {B.}~\bibnamefont {Shao}}, \bibinfo
  {author} {\bibfnamefont {J.-G.}\ \bibnamefont {Li}}, \bibinfo {author}
  {\bibfnamefont {J.}~\bibnamefont {Zou}},\ and\ \bibinfo {author}
  {\bibfnamefont {L.-A.}\ \bibnamefont {Wu}},\ }\href@noop {} {\bibfield
  {journal} {\bibinfo  {journal} {Physical Review A}\ }\textbf {\bibinfo
  {volume} {83}},\ \bibinfo {pages} {052112} (\bibinfo {year}
  {2011})}\BibitemShut {NoStop}%
\bibitem [{\citenamefont {Radhakrishnan}\ \emph {et~al.}(2017)\citenamefont
  {Radhakrishnan}, \citenamefont {Ermakov},\ and\ \citenamefont
  {Byrnes}}]{radhakrishnan2017quantum}%
  \BibitemOpen
  \bibfield  {author} {\bibinfo {author} {\bibfnamefont {C.}~\bibnamefont
  {Radhakrishnan}}, \bibinfo {author} {\bibfnamefont {I.}~\bibnamefont
  {Ermakov}},\ and\ \bibinfo {author} {\bibfnamefont {T.}~\bibnamefont
  {Byrnes}},\ }\href@noop {} {\bibfield  {journal} {\bibinfo  {journal}
  {Physical Review A}\ }\textbf {\bibinfo {volume} {96}},\ \bibinfo {pages}
  {012341} (\bibinfo {year} {2017})}\BibitemShut {NoStop}%
\bibitem [{\citenamefont {Haseli}(2020)}]{Haseli_2020}%
  \BibitemOpen
  \bibfield  {author} {\bibinfo {author} {\bibfnamefont {S.}~\bibnamefont
  {Haseli}},\ }\href {https://doi.org/10.1088/1555-6611/abac65} {\bibfield
  {journal} {\bibinfo  {journal} {Laser Physics}\ }\textbf {\bibinfo {volume}
  {30}},\ \bibinfo {pages} {105203} (\bibinfo {year} {2020})}\BibitemShut
  {NoStop}%
\bibitem [{\citenamefont {Salvia}\ \emph {et~al.}(2023)\citenamefont {Salvia},
  \citenamefont {Mehboudi},\ and\ \citenamefont
  {Perarnau-Llobet}}]{PhysRevLett.130.240803}%
  \BibitemOpen
  \bibfield  {author} {\bibinfo {author} {\bibfnamefont {R.}~\bibnamefont
  {Salvia}}, \bibinfo {author} {\bibfnamefont {M.}~\bibnamefont {Mehboudi}},\
  and\ \bibinfo {author} {\bibfnamefont {M.}~\bibnamefont {Perarnau-Llobet}},\
  }\href {https://doi.org/10.1103/PhysRevLett.130.240803} {\bibfield  {journal}
  {\bibinfo  {journal} {Phys. Rev. Lett.}\ }\textbf {\bibinfo {volume} {130}},\
  \bibinfo {pages} {240803} (\bibinfo {year} {2023})}\BibitemShut {NoStop}%
\bibitem [{\citenamefont {Radaelli}\ \emph {et~al.}(2023)\citenamefont
  {Radaelli}, \citenamefont {Landi}, \citenamefont {Modi},\ and\ \citenamefont
  {Binder}}]{Radaelli_2023}%
  \BibitemOpen
  \bibfield  {author} {\bibinfo {author} {\bibfnamefont {M.}~\bibnamefont
  {Radaelli}}, \bibinfo {author} {\bibfnamefont {G.~T.}\ \bibnamefont {Landi}},
  \bibinfo {author} {\bibfnamefont {K.}~\bibnamefont {Modi}},\ and\ \bibinfo
  {author} {\bibfnamefont {F.~C.}\ \bibnamefont {Binder}},\ }\href
  {https://doi.org/10.1088/1367-2630/acd321} {\bibfield  {journal} {\bibinfo
  {journal} {New Journal of Physics}\ }\textbf {\bibinfo {volume} {25}},\
  \bibinfo {pages} {053037} (\bibinfo {year} {2023})}\BibitemShut {NoStop}%
\bibitem [{\citenamefont {Niezgoda}\ and\ \citenamefont
  {Chwede\ifmmode~\acute{n}\else
  \'{n}\fi{}czuk}(2021)}]{PhysRevLett.126.210506}%
  \BibitemOpen
  \bibfield  {author} {\bibinfo {author} {\bibfnamefont {A.}~\bibnamefont
  {Niezgoda}}\ and\ \bibinfo {author} {\bibfnamefont {J.}~\bibnamefont
  {Chwede\ifmmode~\acute{n}\else \'{n}\fi{}czuk}},\ }\href
  {https://doi.org/10.1103/PhysRevLett.126.210506} {\bibfield  {journal}
  {\bibinfo  {journal} {Phys. Rev. Lett.}\ }\textbf {\bibinfo {volume} {126}},\
  \bibinfo {pages} {210506} (\bibinfo {year} {2021})}\BibitemShut {NoStop}%
\bibitem [{\citenamefont {Lyu}\ \emph {et~al.}(2023)\citenamefont {Lyu},
  \citenamefont {Tang}, \citenamefont {Li}, \citenamefont {Xu}, \citenamefont
  {Yung},\ and\ \citenamefont {Bayat}}]{Lyu_2023}%
  \BibitemOpen
  \bibfield  {author} {\bibinfo {author} {\bibfnamefont {C.}~\bibnamefont
  {Lyu}}, \bibinfo {author} {\bibfnamefont {X.}~\bibnamefont {Tang}}, \bibinfo
  {author} {\bibfnamefont {J.}~\bibnamefont {Li}}, \bibinfo {author}
  {\bibfnamefont {X.}~\bibnamefont {Xu}}, \bibinfo {author} {\bibfnamefont
  {M.-H.}\ \bibnamefont {Yung}},\ and\ \bibinfo {author} {\bibfnamefont
  {A.}~\bibnamefont {Bayat}},\ }\href
  {https://doi.org/10.1088/1367-2630/acd571} {\bibfield  {journal} {\bibinfo
  {journal} {New Journal of Physics}\ }\textbf {\bibinfo {volume} {25}},\
  \bibinfo {pages} {053022} (\bibinfo {year} {2023})}\BibitemShut {NoStop}%
\bibitem [{\citenamefont {Xie}\ \emph {et~al.}(2022)\citenamefont {Xie},
  \citenamefont {Dai}, \citenamefont {Yuan}, \citenamefont {Deng},
  \citenamefont {Li}, \citenamefont {Chen},\ and\ \citenamefont
  {Pan}}]{PhysRevA.106.013316}%
  \BibitemOpen
  \bibfield  {author} {\bibinfo {author} {\bibfnamefont {Y.-J.}\ \bibnamefont
  {Xie}}, \bibinfo {author} {\bibfnamefont {H.-N.}\ \bibnamefont {Dai}},
  \bibinfo {author} {\bibfnamefont {Z.-S.}\ \bibnamefont {Yuan}}, \bibinfo
  {author} {\bibfnamefont {Y.}~\bibnamefont {Deng}}, \bibinfo {author}
  {\bibfnamefont {X.}~\bibnamefont {Li}}, \bibinfo {author} {\bibfnamefont
  {Y.-A.}\ \bibnamefont {Chen}},\ and\ \bibinfo {author} {\bibfnamefont
  {J.-W.}\ \bibnamefont {Pan}},\ }\href
  {https://doi.org/10.1103/PhysRevA.106.013316} {\bibfield  {journal} {\bibinfo
   {journal} {Phys. Rev. A}\ }\textbf {\bibinfo {volume} {106}},\ \bibinfo
  {pages} {013316} (\bibinfo {year} {2022})}\BibitemShut {NoStop}%
\bibitem [{\citenamefont {Mishra}\ and\ \citenamefont {Bayat}(2021)}]{pa1}%
  \BibitemOpen
  \bibfield  {author} {\bibinfo {author} {\bibfnamefont {U.}~\bibnamefont
  {Mishra}}\ and\ \bibinfo {author} {\bibfnamefont {A.}~\bibnamefont {Bayat}},\
  }\href {https://doi.org/10.1103/PhysRevLett.127.080504} {\bibfield  {journal}
  {\bibinfo  {journal} {Phys. Rev. Lett.}\ }\textbf {\bibinfo {volume} {127}},\
  \bibinfo {pages} {080504} (\bibinfo {year} {2021})}\BibitemShut {NoStop}%
\bibitem [{\citenamefont {Montenegro}\ \emph {et~al.}(2022)\citenamefont
  {Montenegro}, \citenamefont {Genoni}, \citenamefont {Bayat},\ and\
  \citenamefont {Paris}}]{pa2}%
  \BibitemOpen
  \bibfield  {author} {\bibinfo {author} {\bibfnamefont {V.}~\bibnamefont
  {Montenegro}}, \bibinfo {author} {\bibfnamefont {M.~G.}\ \bibnamefont
  {Genoni}}, \bibinfo {author} {\bibfnamefont {A.}~\bibnamefont {Bayat}},\ and\
  \bibinfo {author} {\bibfnamefont {M.~G.~A.}\ \bibnamefont {Paris}},\ }\href
  {https://doi.org/10.1103/PhysRevResearch.4.033036} {\bibfield  {journal}
  {\bibinfo  {journal} {Phys. Rev. Res.}\ }\textbf {\bibinfo {volume} {4}},\
  \bibinfo {pages} {033036} (\bibinfo {year} {2022})}\BibitemShut {NoStop}%
\bibitem [{\citenamefont {He}\ \emph {et~al.}(2023)\citenamefont {He},
  \citenamefont {Yousefjani},\ and\ \citenamefont {Bayat}}]{pa3}%
  \BibitemOpen
  \bibfield  {author} {\bibinfo {author} {\bibfnamefont {X.}~\bibnamefont
  {He}}, \bibinfo {author} {\bibfnamefont {R.}~\bibnamefont {Yousefjani}},\
  and\ \bibinfo {author} {\bibfnamefont {A.}~\bibnamefont {Bayat}},\ }\href
  {https://doi.org/10.1103/PhysRevLett.131.010801} {\bibfield  {journal}
  {\bibinfo  {journal} {Phys. Rev. Lett.}\ }\textbf {\bibinfo {volume} {131}},\
  \bibinfo {pages} {010801} (\bibinfo {year} {2023})}\BibitemShut {NoStop}%
\bibitem [{\citenamefont {Lieb}\ \emph {et~al.}(1961)\citenamefont {Lieb},
  \citenamefont {Schultz},\ and\ \citenamefont {Mattis}}]{lieb1961two}%
  \BibitemOpen
  \bibfield  {author} {\bibinfo {author} {\bibfnamefont {E.}~\bibnamefont
  {Lieb}}, \bibinfo {author} {\bibfnamefont {T.}~\bibnamefont {Schultz}},\ and\
  \bibinfo {author} {\bibfnamefont {D.}~\bibnamefont {Mattis}},\ }\href@noop {}
  {\bibfield  {journal} {\bibinfo  {journal} {Annals of Physics}\ }\textbf
  {\bibinfo {volume} {16}},\ \bibinfo {pages} {407} (\bibinfo {year}
  {1961})}\BibitemShut {NoStop}%
\bibitem [{\citenamefont {Wang}\ and\ \citenamefont
  {Zanardi}(2002)}]{wang2002quantum}%
  \BibitemOpen
  \bibfield  {author} {\bibinfo {author} {\bibfnamefont {X.}~\bibnamefont
  {Wang}}\ and\ \bibinfo {author} {\bibfnamefont {P.}~\bibnamefont {Zanardi}},\
  }\href@noop {} {\bibfield  {journal} {\bibinfo  {journal} {Physics Letters
  A}\ }\textbf {\bibinfo {volume} {301}},\ \bibinfo {pages} {1} (\bibinfo
  {year} {2002})}\BibitemShut {NoStop}%
\bibitem [{\citenamefont {Wang}(2002)}]{wang2002thermal}%
  \BibitemOpen
  \bibfield  {author} {\bibinfo {author} {\bibfnamefont {X.}~\bibnamefont
  {Wang}},\ }\href@noop {} {\bibfield  {journal} {\bibinfo  {journal} {Physical
  Review A}\ }\textbf {\bibinfo {volume} {66}},\ \bibinfo {pages} {034302}
  (\bibinfo {year} {2002})}\BibitemShut {NoStop}%
\bibitem [{\citenamefont {Cai}\ \emph {et~al.}(2006)\citenamefont {Cai},
  \citenamefont {Zhou},\ and\ \citenamefont {Guo}}]{cai2006robustness}%
  \BibitemOpen
  \bibfield  {author} {\bibinfo {author} {\bibfnamefont {J.-M.}\ \bibnamefont
  {Cai}}, \bibinfo {author} {\bibfnamefont {Z.-W.}\ \bibnamefont {Zhou}},\ and\
  \bibinfo {author} {\bibfnamefont {G.-C.}\ \bibnamefont {Guo}},\ }\href@noop
  {} {\bibfield  {journal} {\bibinfo  {journal} {Physics Letters A}\ }\textbf
  {\bibinfo {volume} {352}},\ \bibinfo {pages} {196} (\bibinfo {year}
  {2006})}\BibitemShut {NoStop}%
\bibitem [{\citenamefont {Haseli}\ \emph {et~al.}(2020)\citenamefont {Haseli},
  \citenamefont {Haddadi},\ and\ \citenamefont
  {Pourkarimi}}]{haseli2020entropic}%
  \BibitemOpen
  \bibfield  {author} {\bibinfo {author} {\bibfnamefont {S.}~\bibnamefont
  {Haseli}}, \bibinfo {author} {\bibfnamefont {S.}~\bibnamefont {Haddadi}},\
  and\ \bibinfo {author} {\bibfnamefont {M.~R.}\ \bibnamefont {Pourkarimi}},\
  }\href@noop {} {\bibfield  {journal} {\bibinfo  {journal} {Optical and
  Quantum Electronics}\ }\textbf {\bibinfo {volume} {52}},\ \bibinfo {pages}
  {465} (\bibinfo {year} {2020})}\BibitemShut {NoStop}%
\bibitem [{\citenamefont {Ozaydin}\ and\ \citenamefont
  {Altintas}(2015)}]{Ozaydin2015}%
  \BibitemOpen
  \bibfield  {author} {\bibinfo {author} {\bibfnamefont {F.}~\bibnamefont
  {Ozaydin}}\ and\ \bibinfo {author} {\bibfnamefont {A.~A.}\ \bibnamefont
  {Altintas}},\ }\href {https://doi.org/10.1038/srep16360} {\bibfield
  {journal} {\bibinfo  {journal} {Scientific Reports}\ }\textbf {\bibinfo
  {volume} {5}},\ \bibinfo {pages} {16360} (\bibinfo {year}
  {2015})}\BibitemShut {NoStop}%
\bibitem [{\citenamefont {Ben~hammou}\ \emph {et~al.}(2023)\citenamefont
  {Ben~hammou}, \citenamefont {EL~Achab},\ and\ \citenamefont
  {Habiballah}}]{doi:10.1142/S0217979224503211}%
  \BibitemOpen
  \bibfield  {author} {\bibinfo {author} {\bibfnamefont {R.}~\bibnamefont
  {Ben~hammou}}, \bibinfo {author} {\bibfnamefont {A.}~\bibnamefont
  {EL~Achab}},\ and\ \bibinfo {author} {\bibfnamefont {N.}~\bibnamefont
  {Habiballah}},\ }\href {https://doi.org/10.1142/S0217979224503211} {\bibfield
   {journal} {\bibinfo  {journal} {International Journal of Modern Physics B}\
  }\textbf {\bibinfo {volume} {0}},\ \bibinfo {pages} {2450321} (\bibinfo
  {year} {2023})}\BibitemShut {NoStop}%
\bibitem [{\citenamefont {Liu}\ \emph {et~al.}(2020)\citenamefont {Liu},
  \citenamefont {Yuan}, \citenamefont {Lu},\ and\ \citenamefont
  {Wang}}]{liu2020quantum}%
  \BibitemOpen
  \bibfield  {author} {\bibinfo {author} {\bibfnamefont {J.}~\bibnamefont
  {Liu}}, \bibinfo {author} {\bibfnamefont {H.}~\bibnamefont {Yuan}}, \bibinfo
  {author} {\bibfnamefont {X.-M.}\ \bibnamefont {Lu}},\ and\ \bibinfo {author}
  {\bibfnamefont {X.}~\bibnamefont {Wang}},\ }\href@noop {} {\bibfield
  {journal} {\bibinfo  {journal} {Journal of Physics A: Mathematical and
  Theoretical}\ }\textbf {\bibinfo {volume} {53}},\ \bibinfo {pages} {023001}
  (\bibinfo {year} {2020})}\BibitemShut {NoStop}%
\bibitem [{\citenamefont {Helstrom}(1969)}]{helstrom1969quantum}%
  \BibitemOpen
  \bibfield  {author} {\bibinfo {author} {\bibfnamefont {C.~W.}\ \bibnamefont
  {Helstrom}},\ }\href@noop {} {\bibfield  {journal} {\bibinfo  {journal}
  {Journal of Statistical Physics}\ }\textbf {\bibinfo {volume} {1}},\ \bibinfo
  {pages} {231} (\bibinfo {year} {1969})}\BibitemShut {NoStop}%
\bibitem [{\citenamefont {Maroufi}\ \emph {et~al.}(2021)\citenamefont
  {Maroufi}, \citenamefont {Laghmach}, \citenamefont {El~Hadfi},\ and\
  \citenamefont {Daoud}}]{maroufi2021analytical}%
  \BibitemOpen
  \bibfield  {author} {\bibinfo {author} {\bibfnamefont {B.}~\bibnamefont
  {Maroufi}}, \bibinfo {author} {\bibfnamefont {R.}~\bibnamefont {Laghmach}},
  \bibinfo {author} {\bibfnamefont {H.}~\bibnamefont {El~Hadfi}},\ and\
  \bibinfo {author} {\bibfnamefont {M.}~\bibnamefont {Daoud}},\ }\href@noop {}
  {\bibfield  {journal} {\bibinfo  {journal} {International Journal of
  Theoretical Physics}\ }\textbf {\bibinfo {volume} {60}},\ \bibinfo {pages}
  {3103} (\bibinfo {year} {2021})}\BibitemShut {NoStop}%
\bibitem [{\citenamefont {Albarelli}\ \emph {et~al.}(2020)\citenamefont
  {Albarelli}, \citenamefont {Barbieri}, \citenamefont {Genoni},\ and\
  \citenamefont {Gianani}}]{albarelli2020perspective}%
  \BibitemOpen
  \bibfield  {author} {\bibinfo {author} {\bibfnamefont {F.}~\bibnamefont
  {Albarelli}}, \bibinfo {author} {\bibfnamefont {M.}~\bibnamefont {Barbieri}},
  \bibinfo {author} {\bibfnamefont {M.~G.}\ \bibnamefont {Genoni}},\ and\
  \bibinfo {author} {\bibfnamefont {I.}~\bibnamefont {Gianani}},\ }\href@noop
  {} {\bibfield  {journal} {\bibinfo  {journal} {Physics Letters A}\ }\textbf
  {\bibinfo {volume} {384}},\ \bibinfo {pages} {126311} (\bibinfo {year}
  {2020})}\BibitemShut {NoStop}%
\bibitem [{\citenamefont {Carollo}\ \emph {et~al.}(2019)\citenamefont
  {Carollo}, \citenamefont {Spagnolo}, \citenamefont {Dubkov},\ and\
  \citenamefont {Valenti}}]{carollo2019quantumness}%
  \BibitemOpen
  \bibfield  {author} {\bibinfo {author} {\bibfnamefont {A.}~\bibnamefont
  {Carollo}}, \bibinfo {author} {\bibfnamefont {B.}~\bibnamefont {Spagnolo}},
  \bibinfo {author} {\bibfnamefont {A.~A.}\ \bibnamefont {Dubkov}},\ and\
  \bibinfo {author} {\bibfnamefont {D.}~\bibnamefont {Valenti}},\ }\href@noop
  {} {\bibfield  {journal} {\bibinfo  {journal} {Journal of Statistical
  Mechanics: Theory and Experiment}\ }\textbf {\bibinfo {volume} {2019}},\
  \bibinfo {pages} {094010} (\bibinfo {year} {2019})}\BibitemShut {NoStop}%
\bibitem [{\citenamefont {Razavian}\ \emph {et~al.}(2020)\citenamefont
  {Razavian}, \citenamefont {Paris},\ and\ \citenamefont {Genoni}}]{raze20}%
  \BibitemOpen
  \bibfield  {author} {\bibinfo {author} {\bibfnamefont {S.}~\bibnamefont
  {Razavian}}, \bibinfo {author} {\bibfnamefont {M.~G.~A.}\ \bibnamefont
  {Paris}},\ and\ \bibinfo {author} {\bibfnamefont {M.~G.}\ \bibnamefont
  {Genoni}},\ }\bibfield  {journal} {\bibinfo  {journal} {Entropy}\ }\textbf
  {\bibinfo {volume} {22}},\ \href {https://doi.org/10.3390/e22111197}
  {10.3390/e22111197} (\bibinfo {year} {2020})\BibitemShut {NoStop}%
\bibitem [{\citenamefont {Dziarmaga}(2005)}]{dziarmaga2005dynamics}%
  \BibitemOpen
  \bibfield  {author} {\bibinfo {author} {\bibfnamefont {J.}~\bibnamefont
  {Dziarmaga}},\ }\href@noop {} {\bibfield  {journal} {\bibinfo  {journal}
  {Physical review letters}\ }\textbf {\bibinfo {volume} {95}},\ \bibinfo
  {pages} {245701} (\bibinfo {year} {2005})}\BibitemShut {NoStop}%
\bibitem [{\citenamefont {Paris}(2009)}]{paris2009quantum}%
  \BibitemOpen
  \bibfield  {author} {\bibinfo {author} {\bibfnamefont {M.~G.~A.}\
  \bibnamefont {Paris}},\ }\href@noop {} {\bibfield  {journal} {\bibinfo
  {journal} {International Journal of Quantum Information}\ }\textbf {\bibinfo
  {volume} {7}},\ \bibinfo {pages} {125} (\bibinfo {year} {2009})}\BibitemShut
  {NoStop}%
\bibitem [{\citenamefont {Gutenkunst}\ \emph {et~al.}(2007)\citenamefont
  {Gutenkunst}, \citenamefont {Waterfall}, \citenamefont {Casey}, \citenamefont
  {Brown}, \citenamefont {Myers},\ and\ \citenamefont
  {Sethna}}]{gutenkunst2007universally}%
  \BibitemOpen
  \bibfield  {author} {\bibinfo {author} {\bibfnamefont {R.~N.}\ \bibnamefont
  {Gutenkunst}}, \bibinfo {author} {\bibfnamefont {J.~J.}\ \bibnamefont
  {Waterfall}}, \bibinfo {author} {\bibfnamefont {F.~P.}\ \bibnamefont
  {Casey}}, \bibinfo {author} {\bibfnamefont {K.~S.}\ \bibnamefont {Brown}},
  \bibinfo {author} {\bibfnamefont {C.~R.}\ \bibnamefont {Myers}},\ and\
  \bibinfo {author} {\bibfnamefont {J.~P.}\ \bibnamefont {Sethna}},\
  }\href@noop {} {\bibfield  {journal} {\bibinfo  {journal} {PLoS computational
  biology}\ }\textbf {\bibinfo {volume} {3}},\ \bibinfo {pages} {e189}
  (\bibinfo {year} {2007})}\BibitemShut {NoStop}%
\end{thebibliography}%

\end{document}